\documentclass[11pt,draftcls, onecolumn]{IEEEtran}
\usepackage{ifpdf}
\usepackage{cite}
\ifCLASSINFOpdf
 \usepackage[pdftex]{graphicx}
 \DeclareGraphicsExtensions{.pdf,.jpeg,.png}
\else
 \usepackage[dvips]{graphicx}
 \DeclareGraphicsExtensions{.eps}
\fi
\usepackage{amssymb}
\usepackage[cmex10]{amsmath}
\usepackage{array}
\usepackage{mdwmath}
\usepackage{mdwtab}
\usepackage{eqparbox}
\usepackage[tight,footnotesize]{subfigure}
\usepackage{stfloats}

\usepackage{bm}
\usepackage{overpic}
\usepackage{color}
\usepackage{dsfont}
\usepackage{amsthm}
\usepackage{setspace}
\usepackage{mathbbol}
\usepackage[ruled,norelsize]{algorithm2e}
\usepackage{bbm}
\usepackage{boxedminipage}

\newcommand{\E}{\mathds{E}}
\newcommand{\T}{\top}
\newcommand{\mc}{\mathcal}

\renewcommand{\Re}{\mathds{R}}
\newcommand{\St}{\mathds{S}}
\newcommand{\Ac}{\mathds{A}}

\newcommand{\Xf}{\mathds{X}}

\newcommand{\defeq}{\triangleq}
\newcommand{\diag}{\mathrm{diag}}
\newcommand{\N}{\mc{N}}
\newcommand{\col}{\mathrm{col}}

\newcommand{\Pb}{\mathds{P}}

\newtheorem{assumption}{Assumption}

\newtheorem{theorem}{Theorem}

\newcommand{\mysmallarraydecl}{\renewcommand{%
\IEEEeqnarraymathstyle}{\scriptscriptstyle}%
\renewcommand{\IEEEeqnarraytextstyle}{\scriptsize}%
\renewcommand{\baselinestretch}{1.1}%
\settowidth{\normalbaselineskip}{\scriptsize
\hspace{\baselinestretch\baselineskip}}%
\setlength{\baselineskip}{\normalbaselineskip}%
\setlength{\jot}{0.25\normalbaselineskip}%
\setlength{\arraycolsep}{2pt}}
\newlength{\eqboxstorage}

\allowdisplaybreaks 

\title{Distributed Policy Evaluation Under Multiple Behavior Strategies}

\begin{document}

\author{Sergio~Valcarcel~Macua,~\IEEEmembership{Student Member,~IEEE,}
        Jianshu~Chen,~\IEEEmembership{Member,~IEEE}\\
		Santiago~Zazo,~\IEEEmembership{Member,~IEEE,}        
        and~Ali~H.~Sayed,~\IEEEmembership{Fellow,~IEEE}
\thanks{This work was supported in part by the Spanish Ministry of Science
and Innovation in the program CONSOLIDER-INGENIO 2010 under the Grant CSD2008-00010 COMONSENS 
and by the NSF grants CCF-1011918 and ECCS-1407712.
A short preliminary version dealing with a special case of this work appears in the conference publication \cite{Valcarcel2013}.}
\thanks{S. V. Macua and S. Zazo are with the Department of Signals, Systems
and Radiocommunications, Escuela T\'ecnica Superior de Ingenieros de Telecomunicai\'on, 
Universidad Polit\'ecnica de Madrid, Madrid 28040, Spain
(e-mail: sergio@gaps.ssr.upm.es; santiago@gaps.ssr.upm.es).}
\thanks{J. Chen and A. H. Sayed are with the Department of Electrical Engineering, 
University of California, Los Angeles, CA 90095 USA 
(e-mail: cjs09@ucla.edu;  sayed@ee.ucla.edu).}
}


\maketitle

\begin{abstract}
We apply diffusion strategies to develop a fully-distributed cooperative reinforcement learning algorithm 
in which agents in a network communicate only with their immediate neighbors 
to improve predictions about their environment.
The algorithm can also be applied to off-policy learning,
meaning that the agents can predict the response to a behavior
different from the actual policies they are following.
The proposed distributed strategy is efficient, 
with linear complexity in both computation time and memory footprint.
We provide a mean-square-error performance analysis and
establish convergence under constant step-size updates,
which endow the network with continuous learning capabilities.
The results show a clear gain from cooperation:
when the individual agents can estimate the solution,
cooperation increases stability and reduces bias and variance of the prediction error;
but, more importantly, 
the network is able to approach the optimal solution
even when none of the individual agents can 
(e.g., when the individual behavior policies restrict each agent to sample a small portion of the state space).
\end{abstract}


\ifCLASSOPTIONpeerreview
\begin{center} 
	\bfseries EDICS Categories: 
	MLR-DIST; SEN-DCON; SEN-COLB; MLR-COGP 
\end{center}
\fi
%
\IEEEpeerreviewmaketitle

\section{Introduction}
\label{sec:intro}

\IEEEPARstart{C}{onsider} the problem in which a network of autonomous agents
collaborate to predict the response of the environment to their actions. 
The network forms a connected graph, 
where there is at least one path between every pair of nodes.
The agents learn locally from their individual interactions with the environment
and share knowledge with their neighbors. 
Only direct neighborhood communication is allowed.
We assume the environment can be modeled as a Markov decision process.
The agents do not have access to the actual state of the environment, 
but just to feature vectors representing it.
The feature representation is convenient 
in problems with very large state dimensions
since it is computationally more efficient to work with features
of smaller dimension than the size of the original state-space.

In the scenario under study in this work, every agent takes actions according to an individual policy,
which is possibly different from that of every other agent.
The objective of the agents is to assess the response of the environment to a common hypothetical behavior, the \textit{target policy},
which is the same for every agent but different from the actual \emph{behavior policies} they are following.
This problem of predicting the response to a target policy different from the behavior policy is commonly referred as \emph{off-policy} learning \cite{SuttonBarto1998}.
Off-policy learning has been claimed to be necessary when the agents need to perform tasks in complex environments 
because they could perform many different predictions in parallel from a single stream of data \cite{Sutton2011Horde,Degris2012,Modayil2014Nexting}.

The predictions by the agents are made in the form of value functions \cite{Puterman1994,SuttonBarto1998,bertsekas2012dynamic}.
The gradient-temporal-difference (GTD) algorithm is one useful method for computing approximate value functions.
It was originally proposed for the single agent scenario in \cite{Sutton2008,Sutton2009}, 
and derived by means of the stochastic optimization of a suitable cost function. 
The main advantages of this single-agent GTD are its low complexity 
and its convergence guarantees (for diminishing step-sizes)
under the off-policy setting.
In Section \ref{sec:multi-agent} of this work we apply diffusion strategies 
to develop a distributed GTD algorithm that extends the single-agent GTD to multi-agent networks.
There are several distributed strategies that can be used for this purpose, 
such as consensus \cite{Tsitsiklis_1986_DistAsynch, nedic_distributed_2009, Kar2011ConvergenceGossip, Stankovic2011DecentralizedStochasticApprox, Olfati04a} and diffusion strategies \cite{Lopes2008,Cattivelli2010,Chen2012a,Chen2013a_distributed}. 
Consensus strategies have been successfully applied to the solution of static optimization problems, where the objective does not drift with time. 
They have been studied largely under diminishing step-size conditions to ensure agreement among cooperating agents. 
Diffusion strategies, on the other hand, 
have been proved to be particularly apt at endowing networks with continuous adaptation and learning abilities to enable tracking of drifting conditions.
There are several forms of diffusion; recent overviews appear in \cite{SayedChapter2012,Sayed2013SPMagazine,Sayed2014ProcIEEE}. 
It has been shown in \cite{Tu2012} that the dynamics of diffusion networks leads to enhanced stability and lower mean-square-error (MSE) than consensus networks. In particular, the analysis in \cite{Sayed2013SPMagazine, Sayed2014ProcIEEE, Tu2012} shows that
consensus networks combine local data and in-neighborhood information asymmetrically,
which can make the state of consensus networks grow unbounded even when all individual agents are mean stable in isolation. 
This behavior does not happen in diffusion networks, 
in which local and external information are symmetrically combined by construction,
enhancing the stability of the network.
For these reasons, we focus in the remainder of this article on the derivation of a diffusion strategy for GTD over multi-agent networks. 
As a byproduct of this derivation, 
we show that the GTD algorithm,
motivated as a two time-scales stochastic approximation in \cite{Sutton2009},
is indeed a stochastic Arrow-Hurwicz algorithm applied to the dual problem of the original formulation.

The convergence analysis of reinforcement learning algorithms
is usually challenging even for the single-agent case,
and studies are often restricted to the case of diminishing step-sizes
\cite{Sutton2008, Sutton2009, maei_gq_2010}.
For a distributed algorithm, the analysis becomes more demanding
because the estimation process at each node 
is influenced by the estimates at the other nodes, 
so the error propagates across the network.
Another difficulty in the distributed case is that the agents may follow different behavior policies
and, thus, their individual cost functions could have different minimizers.
In Section \ref{sec:mse_analysis}, 
we will analyze the steady-state and transient behavior
of the proposed distributed algorithm,
deriving closed-form expressions that
characterize the network performance
for sufficiently small {\em constant} step-sizes. 
We employ constant, as opposed to decaying step-sizes, because we are interested in distributed solutions that are able to continually adapt and learn.
The performance analysis will reveal that when the agents follow the same behavior policy, 
they will be able to find an unbiased estimator for the centralized solution. 
On the other hand, when the agents behave differently,
they will approach, up to some bias, 
the solution of a convex combination of their individual problems.
This bias is proportional to the step-size, 
so it becomes negligible when the step-size is sufficiently small.
%
%
One important benefit that results when the agents behave differently 
is that, although the agents do not directly share their samples, 
the in-network experience becomes richer in a manner that
the diffusion strategy is able to exploit.
In particular, 
in the reinforcement learning literature,
it is customary to assume that the behavior policy 
must allow the agents to visit every possible state infinitely often. 
We will relax this assumption and show that the distributed algorithm is able to perform well
even when the individual agents only visit small portions of the state-space, 
as long as there are other agents that explore the remaining regions.
Therefore, even though none of the agents can find the optimal estimate of the value function by itself,
they can achieve it through cooperation.
This is an interesting capability that emerges from the networked solution.

In this work, we consider a setting in which the agents can communicate with their neighbors, 
but they operate without influencing each other.
This setup is meaningful in many real applications.
Consider, for example, a water purification plant controlled and monitored by a wireless actuator-sensor network, 
in which each device is attached to a different water-tank.
The quality of the water (e.g., the amount of bacteria) in one tank will be influenced by the decisions (e.g., delivering some amount of chlorine)
made by the device controlling that tank, independently of what other devices do.
Still, since all water tanks behave similarly under similar circumstances, 
the devices in the network can benefit from sharing their individual knowledge.

\subsection{Related works}
\label{ssec:Related_Works}

There are several insightful works in the literature that address issues pertaining to distributed learning albeit under different scenarios and conditions than what is studied in this article. 
For example, the work in \cite{Kar2012} proposes a useful algorithm, 
named QD-learning,
which is a distributed implementation of Q-learning using consensus-based stochastic approximation.
The diffusion strategy proposed herein is different in several respects.
QD-learning asymptotically solves the optimal control problem,
learning the policy that maximizes the long-term reward of the agents.
Here, we focus on predicting the long-term reward for a given policy, 
which is an important part of the control problem.
However, QD-learning is developed in \cite{Kar2012} under the assumption of perfect-knowledge of the state.
Here, we study the case in which the agents only know a feature representation of the state,
which is used to build a parametric approximation of the value function,
allowing us to tackle large problems,
for which Q-learning schemes can diverge \cite{Baird95residualalgorithms,Tsitsiklis1997}.
Finally, we enforce constant step-sizes in order to enable continuous adaptation and learning. 
In comparison, the analysis in \cite{Kar2012} employs a diminishing step-size that dies out as time progresses and, 
therefore, turns off adaptation and is not able to track concept drifts in the data.

Another related work \cite{Bhatnagar2011472} analyzes the performance of cooperative distributed asynchronous estimation 
of linearly approximated value functions using standard temporal difference (TD),
but it is well known that TD learning with parametric approximation schemes can diverge when the agents learn off-policy \cite{Baird95residualalgorithms, Tsitsiklis1997}.
In addition, 
although the algorithm in \cite{Bhatnagar2011472} is distributed, 
in the sense that there is no fusion center, 
it requires full connectivity 
(i.e., every node must be able to exchange information with every other node in the network), 
which is a restrictive assumption that prevents the algorithm from large-scale deployments.
In this article, we focus on fully distributed solutions that only require the network of agents to be connected (but not necessarily fully connected). 
Other related---but more heuristic---approaches include \cite{Schneider99distributedvalue,Varshavskaya08DARS}.

\subsection{Notation}
Lower case letters are used to denote both scalar values and vectors.
Matrices are denoted by upper case letters.
Boldface notation denotes random variables 
(e.g., $s$ is a realization for $\bm{s}$). 
%
The state of the environment and the action taken by an agent are denoted by $s$ and $a$, respectively.
With a slight abuse of notation, $s(i)$ and $a(i)$ denote the state and action variables at time $i$.
Moreover, 
whenever a variable is specific to some agent $k$ we add a subscript
(e.g., $s_k(i) = s$  means that the environment seen by agent $k$ is at state $s$ at time $i$).
%
%

All vectors are column vectors. Superscript $\cdot^\T$ denotes transposition.
The identity matrix of size $S$ is denoted by $I_S$,
the null matrix of size $M\times L$ is denoted by $0_{M\times L}$,
and $\mathbb{1}_M$ and $0_M$ stand for vectors of ones and zeros of length $M$, respectively.
%
The Kronecker product operation is denoted by $\otimes$.
The spectrum, $m$-th eigenvalue and spectral radius of a matrix are denoted by $\lambda(\cdot)$, $\lambda_m(\cdot)$ and $\rho(\cdot)$, respectively.
The operator $\col{\{\cdot\}}$ stacks vectors (or matrices) into a long vector
(or a tall matrix);
while $\text{vec}[\cdot]$ stacks the columns of a matrix, one beneath the other, into a long vector. 
The operator $\diag{\{\cdot\}}$ creates a diagonal matrix (a block-diagonal matrix) from a given vector 
(a set of square matrices).
The Euclidean (semi)norm is given by $\| y \|^2_D \defeq y^\T D y$, 
where $D$ is a positive (semi)definite matrix.
The expected value operator with respect to probability distribution $d$ is denoted by $\E_d [\cdot]$;
we use multiple sub-indexes (e.g., $\E_{d,\phi, \mc{P}} [\cdot]$)
when the expectation is taken with regard to multiple distributions.

\section{Bellman Equation and Value Functions}

\subsection{Markov decision processes (MDP)}
\label{ssec:MDP}

We consider Markov decision processes (MDP) \cite{Puterman1994,bertsekas2012dynamic} that are characterized by 
a finite set of states $\St$ of  size $S \defeq |\St|$;
a finite set of actions $\Ac$;
the kernel of transition probabilities $\mathcal{P}( s' | s, a)$, 
which gives the probability of going 
from one state $s$ to another state $s'$, given an action $a$;
and the reward function $r: \St \times \Ac \times \St \rightarrow \Re$ that the agent wants to predict,
which is associated with every transition,
such that $r \left( s, a, s' \right)$
denotes the reward received by a generic agent
for the transition from $s$ to $s'$
after taking action $a$.

The agents want to predict the response of their environment when they follow some stationary policy $\pi$,
such that $\pi( a | s )$ stands for the probability of an agent choosing action $a$ when the environment is at state $s$.
We assume that the finite-state Markov chain resulting from the MDP is irreducible and aperiodic under any policy of interest.
Thus, it has a unique positive stationary probability distribution of visiting each state \cite[App. A]{Puterman1994} \cite{LevinPeresWilmer2006}
denoted by ${d^\pi} = [d^\pi(1), \ldots, d^\pi(S)]^\T$, such that $d^\pi(s) > 0$, for all $1 \le s \le S $.
The state transition probabilities of the Markov chain, from initial state $s$ to destination $s'$ are given by
\begin{IEEEeqnarray}{rCl}
p_{ss'}^\pi \defeq \Pb \left( s' \: | \: s \right) = \sum_{a\in \Ac} 
	\mathcal{P} \left( s' | s, a \right) \pi (a | s)
\label{eq:P_transition_probabilities}
\end{IEEEeqnarray}
We collect $p_{ss'}^{\pi}$ into an $S \times S$ matrix $P^\pi$ as its $(s,s')$-th entry.

\subsection{Value function}
\label{sec:value-function}

In order to make predictions of the reward signal,
we use state value functions,
$v: \mathcal{\St} \rightarrow \Re$,
which provide the expected cumulative sum of the reward, 
weighted by an exponentially-decaying time window
\cite{Puterman1994, SuttonBarto1998,bertsekas2012dynamic,Modayil2014Nexting}.
This time window spans from $i=0$ to $i=\infty$, 
but it has an effective length 
controlled by a constant $ \gamma \in (0,1)$,
which trades short-sighted ($\gamma \rightarrow 0$) vs. long-term planning ($\gamma \rightarrow 1$).
The value function for target policy $\pi$, starting from some initial state $s \in \St$ at time $i$, is defined as:
\begin{IEEEeqnarray}{rCl}
v^{\pi}(s) \defeq  
		\E_{\pi, \mc{P}} 
			\left[
				\sum_{t=1}^{\infty}
					\gamma^{t-1} 
					\bm{r}(i+t)
			\: 
			\Big| 
			\:
				\bm{s}(i) = s
			\right]
\label{eq:state_val}
\end{IEEEeqnarray}
where
$ \bm{r}(i+1) 	\defeq r ( \bm{s}(i), \bm{a}(i), \bm{s}(i+1) ) $, 
and the expectation is taken with regard to all possible state-transitions.
Note that $\bm{a}(i)$ is random because it is drawn from a probability distribution $\pi$, 
which together with the probabilistic transition dictated by $\mc{P}$, leads to a random future state $\bm{s}(i+1)$.
Let $s'$ denote the destination state after transitioning from $s$.
Then, some algebra will show that we can write \eqref{eq:state_val} as a fixed point equation, 
known as the Bellman equation \cite{Puterman1994, SuttonBarto1998, bertsekas2012dynamic}:
\begin{IEEEeqnarray}{rCl}
v^{\pi} (s) 
&
		=  	
&
			\E_{\pi,\mc{P}} 
				\left[
					\bm{r}(i+1) 
					+ \gamma \bm{r}(i+2)
					+ \ldots
					\: | \:
					\bm{s}(i) = s
				\right]
\nonumber \\
&
=
&
			\E_{\pi, \mc{P}} 
			\left[
				\bm{r} (i+1) 
				\: | \: 
				\bm{s}(i) = s 
			\right]
			+
			\gamma \E_{\pi,\mc{P}} 
				\left[
					\bm{r}(i+2)
					+ \gamma \bm{r}(i+3)
					+ \ldots
					\: | \: 
					\bm{s}(i) = s
				\right]
\nonumber \\
&
=
&
			r^\pi(s)
				+ \gamma \E_{\pi, \mc{P}} 
					\left[
						\sum_{t=1}^{\infty}
						\gamma^{t-1} 
						\bm{r}(i+1+t)
						\: | \: 
						\bm{s}(i) = s
					\right]
\nonumber \\
&
=
&
			r^\pi(s)
				+ \gamma \E_{\pi, \mc{P}} 
					\left[
						v^\pi (s')
							\: | \: 
							\bm{s}(i+1) = s' 
					\right(ght]
\nonumber \\
&
=
&
			r^\pi(s)
				+ \gamma \sum_{s'\in \St} 
							\sum_{a\in \Ac} 
								\mathcal{P} 
									\left( 
										s' | s, a 
									\right) 
									\pi (a | s)
						v^\pi (s')
\nonumber \\
&
=
&
			r^\pi(s)
				+ \gamma \sum_{s'\in \St} p_{ss'}^\pi
						v^\pi (s')
\label{eq:bellman-equation-expansion}
\end{IEEEeqnarray}
where $r^\pi(s)$ denotes the expected reward 
that can be collected over the next transition when the agent is currently at state $s$:
\begin{IEEEeqnarray}{rCl}
	r^\pi (s) 
		& \defeq & 
			\E_{\pi, \mc{P}} [ r (s, \bm{a}, \bm{s}') ]
		=
			\sum_{a \in \Ac} \pi(a | s) 
			\sum_{s' \in \St} \mc{P}(s' | s,a) r (s,a,s')		
\label{eq:expected-reward}
\end{IEEEeqnarray}
Let $v^\pi$ and $r^\pi$ be the vectors of length $S$ that collect 
the values $v^\pi(s)$ and $r^\pi(s)$ for all $s \in \St$, respectively:
\begin{IEEEeqnarray}{rCl}
\label{eq:vector-values-features}
	v^\pi \defeq 
		\left[ 
			{\begin{array}{c} 
				v^\pi(1)
				\\ 
				\vdots
				\\
				v^\pi(S)
			\end{array}} 
		\right]
	\in \Re^S		
		,
		\quad
	r^\pi \defeq 
		\left[ 
			{\begin{array}{c} 
				r^\pi(1)
				\\ 
				\vdots
				\\
				r^\pi(S)
			\end{array}} 
		\right]
	\in \Re^S
\end{IEEEeqnarray}
Then, Eq. \eqref{eq:bellman-equation-expansion} can be written in vector form as the linear system of equations:
\begin{IEEEeqnarray}{rCl}
	(I_S - \gamma P^\pi ) v^\pi  & = & 	r^\pi
\label{eq:linear-Bellman-equation}
\end{IEEEeqnarray}
We shall refer to $v^{\pi}$ as the value vector.
There are two challenges when we aim to obtain $v^\pi$ from \eqref{eq:linear-Bellman-equation}.
The first challenge is that the size of the state-space can be very large
(e.g., the \textit{chess} game has $10^{47}$ possible states,
making \eqref{eq:linear-Bellman-equation} computationally intractable).
The second challenge arises when the agents do not know anything about the environment, 
thus $P^\pi$ and $r^\pi$ are unavailable.
In the following subsections we review how to address these two issues.

\subsection{Approximate value function as a saddle-point problem}
\label{ssec:Approximate-value-function}

For the single agent scenario, 
references \cite{Sutton2008,Sutton2009} introduced efficient algorithms with convergence guarantees under general conditions.
The algorithms save on computations by relying on features that span a space of much lower dimensionality than the size of the original state space.
More formally, 
let $x: \St \rightarrow \Re^M$ be some mapping from states to features,
such that $x_s$ is the \emph{feature vector} of length $M \ll S$
that represents the state $s$.
Now, it would be efficient to approximate the original value function $v^{\pi}(s)$  
as a parametric function of $x_s$,
for some parameter vector $w \in \Re^M$.
When this is done, the problem of making a prediction (i.e., estimating the value vector $v^\pi$)
becomes equivalent to seeking a parameter vector $w^{\star}$ 
that is optimal in a certain sense.
Among many parametrizations, 
a linear approximation of the form
\begin{IEEEeqnarray}{rCl}
\label{eq:approx-val} 
	v^\pi(s) 	\approx  	x_s^\T w
\end{IEEEeqnarray}
has been extensively studied in the literature
(see, e.g., \cite{Tsitsiklis1997, Scherrer2010, Geist2011})
and it is promising mainly because it leads to solutions with low computational demands. 
Moreover, it is expected that if one chooses the mapping of features carefully, 
then the linear approximation model will generally provide good results 
(see, e.g., 
\cite{
	Menache05basisfunction, 
	Parr2008, 
	yu2009basis, 
	mahadevan2009learning, 
	Boots11a,
	bellemare2012sketches
	}
).
Let $X$ be the matrix of size $S \times M$
formed by stacking the transposed feature vectors, $x_s^\T$, on top of each other:
\begin{IEEEeqnarray}{rCl}
\label{eq:features}
	X 	& \defeq &
				\left[ 
				{\begin{array}{c} 
					x_1^\T
					\\
					\vdots
					\\
					x_S^\T
				\end{array}} 
			\right] 		
			\in \Re^{S \times M}
\end{IEEEeqnarray}
Then, the linear approximation 
\eqref{eq:approx-val} 
can be expressed in vector form as:
\begin{IEEEeqnarray}{rCl}
\label{eq:vector-approx-val} 
	v^\pi 	& \approx & 		X w
\end{IEEEeqnarray}
By modeling the value function in the form \eqref{eq:vector-approx-val},
we solve for $w$ by using the approximation \eqref{eq:vector-approx-val} 
in \eqref{eq:linear-Bellman-equation}.
Doing so leads to the approximate Bellman equation:
\begin{IEEEeqnarray}{rCl}
\label{eq:vector-fixed-point}
	Xw  		& = & 	r^\pi + \gamma P^\pi Xw
\label{eq:approximated-vector-Bellman}
\end{IEEEeqnarray}
In this paper, we assume that the features available for the agents constitute a linearly independent set of basis functions,
which effectively represent the states.
Thus, $X$ is {\em full rank} by construction.
However, the fixed point equation \eqref{eq:approximated-vector-Bellman} may not have a solution $w$ in general 
because the right-hand side need not lie in the range space of $X$, which we denote by $\Xf$. 
To address this issue, one approach is to solve instead the \textit{projected Bellman equation} \cite{Tsitsiklis1997}:
\begin{IEEEeqnarray}{rCl}
\label{eq:projected-vector-fixed-point}
	Xw 		&=&	 	\Pi( r^\pi + \gamma P^\pi Xw )
\end{IEEEeqnarray}
where $\Pi$ is a projection operator onto $\Xf$. 
Since $\Xf$ is a linear space, the projection operator with respect to some metric norm
$\| \cdot \|_{D}$ is defined as:
\begin{IEEEeqnarray}{rCl}
	\Pi x 
			\defeq \arg\min_{\bar{x} \in \Xf} \| x - \bar{x} \|_D^2
\end{IEEEeqnarray}
where $D$ is a symmetric positive-definite matrix.
The matrix $\Pi$ is given by
\begin{IEEEeqnarray}{rCl}
\label{eq:projector}
	\Pi 	&=&	 	X(X^\T D X)^{-1} X^\T D
\end{IEEEeqnarray}
Therefore, for different choices of $D$, 
we have different projection operators. 
However, some choices for $D$ will lead to simpler solutions, as we will reveal in Subsection \ref{ssec:primal-dual-approach}. 

Equation \eqref{eq:projected-vector-fixed-point} is now an over-determined consistent  linear system of equations. To solve for $w$, 
reference \cite{Sutton2009} considered the weighted least-squares problem:
\begin{IEEEeqnarray}{rCl}
\label{eq:projected-Bellman-error}
	\underset{w}{\rm minimize} \quad	&
		J_{\rm PB} (w) 	\defeq 	\| \Pi ( r^\pi + \gamma P^\pi Xw ) - Xw \|_D^2
		\quad
\end{IEEEeqnarray}
where the cost function $J_{\text{PB}} (w)$ is referred to as the \textit{projected Bellman error}.
Since $Xw$ already lies in $\Xf$ and $D$ is positive definite,
it can be verified that 
\begin{IEEEeqnarray}{rCl}
	J_{\rm PB}
		(w)		
&
		=
&		
			\| 
				\Pi r^{\pi} - \Pi(I_S - \gamma P^{\pi})Xw
			\|^2_{D}
\nonumber	\\
&
		= 
&
			\left(
				r^{\pi} - (I_S - \gamma P^{\pi})Xw
			\right)^\T
			\Pi^\T D \Pi
			\left(
				r^{\pi} - (I_S - \gamma P^{\pi})Xw
			\right)
\nonumber	\\
&
		= 
&
			\left(
				X^\T D r^{\pi} - B w
			\right)^\T
			\left(
				X^\T D X
			\right)^{-1}
			\left(
				X^\T D r^{\pi} - B w
			\right)				
\label{eq:Projected-Bellman-Least-Squares}
\end{IEEEeqnarray}
where
$
	B 
	\defeq 
		X^\T D (I_S - \gamma P^{\pi})X
$.
Using \eqref{eq:Projected-Bellman-Least-Squares}, 
it can also be verified that the solution $w^\star$ that minimizes $J_{\rm PB}(w)$ satisfies the following normal equations \cite{Sayed2008}:
\begin{IEEEeqnarray}{rCl}
\label{eq:normal-equations}
	B^\T (X^\T D X)^{-1} B w^\star 		& = &		B^\T (X^\T D X )^{-1} X^\T D r^{\pi}  
\end{IEEEeqnarray}
Since $\| P^\pi \|_\infty = 1$ and $\gamma < 1$, 
we can bound the spectral radius of $\gamma P^\pi $ by
\begin{IEEEeqnarray}{rCl}
\label{eq:spectral-radius-gammaP}
	\rho(\gamma P^\pi ) \le \| \gamma P^\pi \|_\infty  = \gamma < 1
\end{IEEEeqnarray}
Thus, the inverse 
$( I_S - \gamma P^\pi )^{-1}$ exists.
In addition, since the matrices $D$ and $X$ have full-rank by assumption,
we conclude that
matrix $B$ is invertible, 
so the minimizer $w^\star$ is given by 
\begin{IEEEeqnarray}{rCl}
	w^\star 	& = &	\left(
						X^\T D 
						\left(
							I_S -\gamma P^{\pi}
						\right)
						X
					\right)^{-1} 
					X^\T D r^{\pi} 
\label{eq:optimal-w}
\end{IEEEeqnarray}
If the quantities $\{P^{\pi}, r^{\pi}\}$ were known, 
one would proceed to solve \eqref{eq:optimal-w} 
and determine the desired vector $w^\star$ and the sought-after value vector $v^{\pi}$ from \eqref{eq:vector-approx-val}.  
However, we want the agents to learn $v^{\pi}$ without any prior knowledge of the environment. 
In other words, we cannot assume $P^{\pi}$ and $r^{\pi}$ are known. 
For this reason, we need to develop an alternative solution method. 
In the process of doing so, first for single-agents, we shall arrive at the same gradient temporal difference method of \cite{Sutton2009} 
albeit by using a fundamentally different approach involving a primal-dual argument. 
The approach will subsequently enable us to generalize to a fully distributed solution. 

So let us continue with the single-agent case for now. 
Our first step relies on relating Eq. \eqref{eq:Projected-Bellman-Least-Squares} to the saddle-point conditions of a convex optimization problem.  
Indeed, minimizing $J_{\rm PB} (w)$ in \eqref{eq:projected-Bellman-error} is equivalent to the following quadratic programming problem:
\begin{IEEEeqnarray}{rCl}
	\begin{aligned}
		\underset{\varepsilon, w}{\rm minimize} 	&\quad 	\frac{1}{2} \varepsilon^\T ( X^\T D X )^{-1} \varepsilon 
\\
		{\rm s.t.} 									&\quad 	\varepsilon = X^\T D r^{\pi} - B w
	\end{aligned}
\label{eq:primal-problem}
\end{IEEEeqnarray}
where we have introduced the splitting variable $\varepsilon$. 
Since problem \eqref{eq:primal-problem} is convex and satisfies Slater's condition \cite{boyd2004convex},
strong duality holds and the primal and dual optimal values are attained and equal 
and they form a saddle-point of the Lagrangian.
Specifically, the Lagrangian of \eqref{eq:primal-problem} is 
\begin{IEEEeqnarray}{rcl}
	L(\varepsilon, w, \theta) 
		& = & \frac{1}{2} \| \varepsilon \|^2_{( X^\T D X )^{-1}}
				+ \theta^\T 
				\left( 
					X^\T D r^{\pi} 
					- B w - \varepsilon
				\right)	
\quad \;\;
\label{eq:primal-lagrangian}
\end{IEEEeqnarray}
where $\theta$ is the Lagrange multiplier.
By minimizing $L(\varepsilon, w, \theta)$	over $\varepsilon$ and $w$, 
we obtain that the dual function is $g(\theta) = - \infty$ unless $B^\T \theta = 0_M$, 
	in which case we have
	\begin{IEEEeqnarray}{rCl}
		g(\theta) 
				 =
					- \frac{1}{2} \theta^\T {X^\T D X} \theta
					+ \theta^\T  X^\T D r^{\pi} 
	\label{eq:dual-function}
	\end{IEEEeqnarray}
	Therefore, the dual problem of \eqref{eq:primal-problem} is given by
	\begin{IEEEeqnarray}{rCl}
	\begin{aligned}
	\underset{\theta}{\rm minimize}	& \quad 	
												\frac{1}{2} \theta^\T X^\T D X \theta
												- \theta^\T X^\T D r^\pi
	\\
		{\rm s.t.} 					&\quad		B^\T \theta = 0_M
	\end{aligned}
	\label{eq:dual-problem}
	\end{IEEEeqnarray}
The main reason to solve \eqref{eq:dual-problem} 
instead of the primal problem \eqref{eq:primal-problem} 
is that the dual formulation removes the inverse in the weighting matrix, $X^\T D X$.
This transformation brings two benefits.
First, in Sec. \ref{ssec:primal-dual-approach}, we will see that it is straightforward to optimize \eqref{eq:dual-problem} from samples.
Second,
as it is explained in Sections \ref{sec:multi-agent} and \ref{sec:existence-uniqueness-solution},
problem  \eqref{eq:dual-problem} leads to a distributed algorithm
in which the agents are able to combine their individual experience to solve the problem.

Had we assumed $P^\pi$ and $r^\pi$ to be known, 
problem \eqref{eq:dual-problem} would be trivial, with unique solution $\theta = 0_M$.
However, since we do not assume any prior knowledge, 
we are going to employ instead a primal-dual algorithm 
that leads to an iterative stochastic-approximation mechanism to learn from samples. 
First, we derive the Lagrangian of \eqref{eq:dual-problem} as
\begin{IEEEeqnarray}{rCl}
	L(\theta, w)  	
		& = &	
			\frac{1}{2} \theta^\T X^\T D X \theta 
			- \theta^\T X^\T D r^\pi 
			+ w^\T B^\T \theta
\nonumber \\
		& = &
			\theta^\T X^\T D 
				\left(
				\frac{1}{2} 
					X \theta 
					+ (I_S - \gamma P^\pi) X w 
					- r^\pi 
				\right)
\label{eq:Lagrangian}
\end{IEEEeqnarray}
where $w$ denotes the Lagrange multiplier.
We use the same notation $w$ to denote the dual variable for \eqref{eq:Lagrangian} because 
it can be verified that by computing the dual of the dual problem \eqref{eq:dual-problem} we recover the original problem \eqref{eq:projected-Bellman-error},
which is equivalent to \eqref{eq:primal-problem}.
Thus, the optimal dual variable $w^\star$ of \eqref{eq:Lagrangian} is also the optimal solution to \eqref{eq:projected-Bellman-error}.
To find a saddle-point $\{ \theta^\star, w^\star \}$ of the Lagrangian \eqref{eq:Lagrangian}
we alternate between applying gradient descent to $L(\theta,w)$ with respect to $\theta$ 
and gradient ascent with respect to $w$:
\begin{IEEEeqnarray}{rCl}
\IEEEnosubnumber
	\theta_{i+1} 		& = &		\theta_i - \mu_\theta X^\T D 
																\left( 
																	X \theta_i + 
																	(I_{S} - \gamma P^\pi) X w_i 
																	- r^\pi 
																\right) 
\qquad
\IEEEyessubnumber \label{eq:dual-fixed-point}\\
	w_{i+1} 			& = &		w_i + \mu_w X^\T(I_S-\gamma P^{\pi})^\T D X \theta_i
\IEEEyessubnumber \label{eq:primal-fixed-point}
\end{IEEEeqnarray}
where $\mu_\theta$ and $\mu_w$ are positive step-sizes.

Construction \eqref{eq:dual-fixed-point}--\eqref{eq:primal-fixed-point} is the well-known Arrow-Hurwicz algorithm 
(see, e.g., \cite{Benzi2005}, \cite[Ch. 9.3.3]{poliak1987introduction} and \cite[Ch. 10]{arrow1958studies}).


\subsection{Primal-dual stochastic optimization}
\label{ssec:primal-dual-approach}

As mentioned before, since the agents do not have prior knowledge of the environment,
we need to replace \eqref{eq:dual-fixed-point}--\eqref{eq:primal-fixed-point} by constructions that do not depend on the quantities $\{P^{\pi},r^{\pi}\}$. 
In order to find the solution directly from samples, we need to convert these gradient iterations into stochastic approximations.
The selection of an appropriate weighted norm $\| \cdot \|_{D}$ in \eqref{eq:projected-Bellman-error} now becomes relevant.
If we choose a weighting matrix $D$ that represents the probability distribution of visiting each state, 
then we can express the terms that appear in \eqref{eq:dual-fixed-point}--\eqref{eq:primal-fixed-point} 
as expectations that we can substitute with their sample estimates.
We proceed to explain the details.


Let us set the weighting matrix in \eqref{eq:projected-Bellman-error} equal to the state-visitation probability induced by the behavior policy
(which we emphasize with the corresponding superscript), i.e., 
$
	D 		\defeq 		D^\phi  		\defeq 	\diag \{ d^\phi \}
$.
Equations \eqref{eq:dual-fixed-point}--\eqref{eq:primal-fixed-point}
depend on  $P^\pi$ and  $r^\pi$, 
meaning that the agent aims to predict the value vector along the expected trajectory 
that would have been induced by the target policy $\pi$.
However, 
the state-visitation distribution of this trajectory, $d^\pi$,
does not match the distribution of the samples actually gathered by the agent, given by $d^\phi$.
Importance sampling \cite[Ch. 9.7]{kroese_handbook_2011} is a technique for estimating properties of a particular distribution, 
while only having samples generated from a different distribution.
Let us introduce \textit{importance weights} 
that measure the dissimilarity between the target ($\pi$) and behavior ($\phi$) policies.
\begin{IEEEeqnarray}{rCl}
	\xi (a, s) 		& \defeq &		\frac{\pi(a|s)}{\phi(a|s)}
\label{eq:importance-weights}
\end{IEEEeqnarray}
By using importance sampling, reference \cite{maei_gq_2010} showed that we can write the gradient inside \eqref{eq:dual-fixed-point}
in terms of moment values of the behavior policy as follows:
\begin{IEEEeqnarray}{rCl}
X^\T
&&
	D^{\phi} 
	\left( 
		X \theta_i 
		+ 
		(I_{S} - \gamma P^\pi) X w_i 
		- r^\pi 
	\right) 
\nonumber\\
	&& 
	\;
	= 
	\:
	\sum_{s \in \St} 
		d^\phi(s) 
		x_s
		\Bigg(
			x_s^\T \theta_i
			+
			\left(
				x_s^\T
				-
				\gamma
				\sum_{s' \in \St} 
					p^\pi_{ss'} 
					x_{s'}^\T
			\right)
			w_i
			-
			\sum_{a \in \Ac}
			\sum_{s' \in \St}
				\mc{P}(s'|s,a)
				\pi(a|s) 
				r(s,a,s')
		\Bigg)
\nonumber \\
		&& 
		\;
		= \:
			\sum_{s \in \St} 
				\sum_{a \in \Ac}
					\sum_{s' \in \St}
						\mc{P}(s'|s,a)
						\pi(a|s)		
						d^\phi(s) 
		\cdot
			x_s
			\left(
				x_s^\T \theta_i
				+
				(
					x_s
					-
					\gamma x_{s'}
				)^\T
				w_i
				-
				r(s,a,s')
			\right)
\nonumber \\
		&& 
		\;
		= \:
			\sum_{s \in \St} 
				\sum_{a \in \Ac}
					\sum_{s' \in \St}
						\mc{P}(s'|s,a)
						\phi(a|s)
						\xi(a,s)		
						d^\phi(s) 						
		\cdot
			x_s
			\left(
				x_s^\T \theta_i
				+
				(
					x_s
					-
					\gamma x_{s'}
				)^\T
				w_i
				-
				r(s,a,s')
			\right)
\nonumber \\
		&& 
		\;
		= 
			\E_{d^\phi, \phi, \mc{P}} 
				\big[
					\bm{x}_s 
					\big(
						\bm{x}_s^\T 
						\theta_i
						+
						(
							\bm{x}_s	
							-
							\gamma \bm{x}_{s'}
						)^\T
						w_i
						-
						r(\bm{s},\bm{a},\bm{s}')
					\big)
					\xi(\bm{a},\bm{s})
				\big] 
\label{eq:theta-moment-values}
\end{IEEEeqnarray}
Similarly, we can express the gradient inside \eqref{eq:primal-fixed-point} as
\begin{IEEEeqnarray}{rCl}
X^\T (I_{S} - 
\gamma P^\pi)^\T D^{\phi} X \theta_i 
& = &
			\E_{d^\phi, \phi, \mc{P}} 
				\left[
					(
						\bm{x}_s	
						-
						\gamma \bm{x}_{s'}
					)
					\bm{x}_s^\T
					\xi(\bm{a},\bm{s})
				\right] 	
				\theta_i
\label{eq:w-moment-values}
\end{IEEEeqnarray}
The agent does not know these expected values though.
Rather, at every time-step,
the agent observes transitions of the form $\{x_i,a(i),x_{i+1},r(i+1)\}$,
where $x_i \defeq x_{s(i)}$ denotes the feature vector observed at time $i$.

In addition, the agent knows both its behavior policy $\phi$
and the target policy $\pi$ that it wants to evaluate
so it can compute the importance weight.
Nevertheless, in an actual implementation, 
the agent need not know the states but just features,
hence, the actual policies must be conditioned on the feature vectors.
Slightly abusing notation, we introduce the importance weight that the node 
computes at every time step:
\begin{IEEEeqnarray}{rCl}
	\xi(i) 				& \defeq & 		\frac{\pi(a(i)|x_i)}{\phi(a(i)|x_i)}
						\approx			
										\frac{\pi(a(i)|s(i))}{\phi(a(i)|s(i))}
						\defeq			\xi(a(i), s(i))
\;\;
\label{eq:importance-weight-from-features}
\end{IEEEeqnarray}
Since a sample of a random variable is an unbiased estimator of its expected value, 
we can build a pair of stochastic approximation recursions from 
\eqref{eq:dual-fixed-point}--\eqref{eq:primal-fixed-point} and \eqref{eq:theta-moment-values}--\eqref{eq:w-moment-values}:
\begin{IEEEeqnarray}{rCl}
\IEEEnosubnumber
	\theta_{i+1} 		& = &	\theta_{i} 
								- \mu_\theta  
									x_{i}
									\left( 
										 x_{i}^\T \theta_{i}
										+ 
										\delta_{i+1}^\T w_i 
										- r(i+1)
									\right)
									\xi(i) 
\quad\quad
\IEEEyessubnumber \label{eq:dual-stochastic-approx} \\
	w_{i+1} 			& =	&	w_{i}  + \mu_w \delta_{i+1} x_i^\T \theta_{i}\xi(i)
\IEEEyessubnumber \label{eq:primal-stochastic-approx}
\end{IEEEeqnarray}
where we introduced 
$	\delta_{i+1} \defeq  x_i - \gamma x_{i+1} $.
Recursions \eqref{eq:dual-stochastic-approx}--\eqref{eq:primal-stochastic-approx} coincide with the single-agent gradient-temporal difference (GTD2) algorithm, which was derived in \cite{Sutton2009} using a different approach. 
The above derivation from \eqref{eq:primal-problem} to \eqref{eq:primal-stochastic-approx} shows that GTD2 is a
stochastic Arrow-Hurwicz algorithm applied to the dual problem of \eqref{eq:projected-Bellman-error}.
More importantly, 
as we will see in the following sections, 
the primal-dual approach is convenient for a multi-agent formulation, 
since it leads to a meaningful in-network state-visitation distribution 
that combines the individual stationary distributions of the agents, 
thus overcoming non-exploratory individual behavior policies. 
\section{Multi-Agent Learning}
\label{sec:multi-agent}

We now consider a network of $N$ connected agents that operate in similar but independent MDPs.
The state-space $\St$, action-space $\Ac$, 
and transition probabilities $\mc{P}$ are the same for every node,
but their actions do not influence each other.
Thus, the transition probabilities seen by each agent $k$
are only determined by its own actions,
$a_k(i) \in \Ac$, 
and the previous state of its environment,
 $s_k(i) \in \St$:
\begin{IEEEeqnarray}{rCl}
	s_k(i+1) \sim \mc{P} (\cdot|s_k(i), a_k(i)), \quad k = 1 \ldots N
\label{eq:independent-MDPs}
\end{IEEEeqnarray}
This assumption is convenient because it makes the problem stationary 
without forcing each agent to know the actions and feature vectors of every other agent in the network.
The agents aim to predict the response of their environment to a common \textit{target} policy $\pi$ 
while they follow different \textit{behavior} policies, 
denoted by $\phi_k(a|s)$ each.

Motivated by recent results on network behavior in \cite{Chen2012b, Sayed2014ProcIEEE}, 
we note that, through collaboration, each agent may contribute to the network with its own experience.
Let $D^{\phi_k}$ be the diagonal matrix that represents the 
stationary state-visitation distribution for agent $k$.
We then introduce the following global problem in place of \eqref{eq:dual-problem} with $D$ substituted by $D^{\phi_k}$:
\begin{IEEEeqnarray}{rCl}
\begin{aligned}
	\underset{\theta}{\rm minimize} 	
		&\quad 
						\sum_{k=1}^N 
							\tau_k
							\left( 
								\frac{1}{2} \theta^\T X^\T D^{\phi_k} X \theta
								- \theta^\T X^\T D^{\phi_k} r^\pi
							\right)
\\
	{\rm s.t.} 		&\quad
						\sum_{k=1}^N \tau_k	
											\big(
												X^\T D^{\phi_k} 
												(I_S - \gamma P^{\pi})X 
											\big)^\T 
											\theta = 0
\end{aligned}
\label{eq:global-dual-problem}
\end{IEEEeqnarray}
where 
$\tau = [\tau_1, \ldots, \tau_N]^\T$ is a vector of non-negative parameters 
whose purpose is to weight the contribution of each agent's local problem to the global problem, such that $\tau^\T \mathds{1}_N = 1$.
Since the dual problem \eqref{eq:dual-problem} removes the inverse of the weighting matrices $X^\T D^{\phi_k} X$, 
we can introduce the in-network stationary distribution
\begin{IEEEeqnarray}{rCl}
\label{eq:global-stationary-distribution}
	D^{\overline{\phi}} 
		& \defeq &
			\sum_{k=1}^N \tau_k  D^{\phi_k}
\end{IEEEeqnarray}
Note that solving the aggregated problem \eqref{eq:global-dual-problem} is effectively solving the single-agent problem \eqref{eq:dual-problem} with $D$ replaced by $D^{\overline{\phi}}$.
The Lagrangian of \eqref{eq:global-dual-problem} is given by
\begin{IEEEeqnarray}{rCl}
	L(\theta, w)  	  & = & 
							\sum_{k=1}^N 
								\tau_k
								L_k(\theta, w) 
\label{eq:global-Lagrangian}
\end{IEEEeqnarray}
where the individual Lagrangians are given by 
\begin{IEEEeqnarray}{rCl}
	L_k(\theta, w)  	  & = & 
							\theta^\T X^\T D^{\phi_k}
							\left(
								\frac{1}{2}  X \theta 
								+ (I_S - \gamma P^\pi) X w 
								- r^\pi 
							\right)	
\quad\;\;
\label{eq:individual-Lagrangian}
\end{IEEEeqnarray}
which are similar to \eqref{eq:Lagrangian} but with stationary distribution $D^{\phi_k}$.
In order to find the global saddle-point of the aggregate Lagrangian \eqref{eq:global-Lagrangian} 
in a cooperative and stochastic manner,
we apply diffusion strategies \cite{SayedChapter2012, Sayed2013SPMagazine, Sayed2014ProcIEEE}.
We choose the \textit{adapt-then-combine} (ATC) diffusion variant 
for distributed optimization over networks \cite{Chen2012a,Chen2012b,Chen2013a_distributed}.
The algorithm consists of two-steps:
the \textit{adaptation} step,  
at which every agent updates its own intermediate estimate independently of the other agents;
and the \textit{combination} step, 
at which every agent combines its neighbors' estimates.
Similar to the derivation of the single-agent algorithm 
\eqref{eq:dual-stochastic-approx}--\eqref{eq:primal-stochastic-approx},
we can express the gradient of the individual Lagrangians \eqref{eq:individual-Lagrangian} in terms of moment values
(i.e., replacing $D^{\phi}$ by $D^{\phi_k}$ into \eqref{eq:theta-moment-values}--\eqref{eq:w-moment-values}). 
We then follow a primal-dual approach and apply 
ATC
twice:
{\em i)} for minimizing $L_k(\theta, w)$ in \eqref{eq:individual-Lagrangian} 
over $\theta$ through stochastic gradient descent:
\begin{IEEEeqnarray}{rcl}
\IEEEnosubnumber
	\widehat{\theta}_{k,i+1} 	
		& = &
			\theta_{k,i} 
			- 
			\mu_\theta  
				x_{k,i}
				\big( 
					 x_{k,i}^\T \theta_{k,i}
					+ 
					\delta_{k,i+1}^\T w_{k,i} 
					-
					r_k(i+1)					
				\big) 
				\xi_k(i)
\notag\\
\IEEEyessubnumber \label{eq:odgtd-adapt-theta} \\
	\theta_{k,i+1} 	
		& = & 
			\sum_{l\in\N_k} c_{lk} \: \widehat{\theta}_{l,i+1} 
\IEEEyessubnumber \label{eq:odgtd-combine-theta}
\end{IEEEeqnarray}
and {\em ii)} for maximizing $L_k(\theta, w)$ in \eqref{eq:individual-Lagrangian} over $w$ through stochastic gradient ascent:
\begin{IEEEeqnarray}{rCl}
	\widehat{w}_{k,i+1} 	
		& = & 
			w_{k,i} 
			+ \mu_w 
				\delta_{k,i+1}  x_{k,i}^\T \theta_{k,i}
				\xi_k(i)
\IEEEyessubnumber \label{eq:odgtd-adapt-w} \\
	w_{k,i+1} 	
		& = & 
			\sum_{l\in\N_k} c_{lk} \: \widehat{w}_{l,i+1} 
\IEEEyessubnumber \label{eq:odgtd-combine-w}
\end{IEEEeqnarray}
where $\N_k$ stands for the neighborhood of agent $k$ 
(i.e., the set of agents that are able to communicate with agent $k$ in a single hop, including $k$ itself),
$\hat{\theta}$ and $\hat{w}$ correspond to the locally adapted estimates,
and $\theta$ and $w$ correspond to the combined estimates for the adapt-then-combine strategy.
The combination coefficients $\{ c_{lk} \}$ define the weights on the links in the network
and can be chosen freely by the designer, 
as long as they satisfy:
\begin{IEEEeqnarray}{rCl}
	c_{lk}  & \ge & 0
, \;\;\;
	\sum_{l\in \N_k} c_{lk}  = 1
, \;\;\;
	c_{lk}  = 	0 \:\; \text{if} \:\; l \notin \N_k	
\label{eq:Comb_matrix_1_2_3}
\\
	c_{kk}  & > & 0  \:\; \text{for at least one agent } k
\label{eq:Comb_matrix_4}
\end{IEEEeqnarray}
Let $C  \defeq [ c_{lk} ]$ be the combination matrix.
Then, condition \eqref{eq:Comb_matrix_1_2_3}
implies that $C$ 
is left-stochastic.
Condition \eqref{eq:Comb_matrix_4} means that 
there is at least one agent that trusts its local measurements and is able to perform its own adaptation step.
We also assume that the topology of the network is connected
(i.e., there is at least one path between any pair of nodes)
and that the combination matrix $C$ remains fixed over time.
Therefore, conditions \eqref{eq:Comb_matrix_1_2_3}--\eqref{eq:Comb_matrix_4} ensure that $C$ is a primitive matrix 
(i.e., there exists $j > 0$ such that all entries of $C^j$ are strictly positive) \cite{SayedChapter2012, seneta2006non}.
It follows from the Perron-Frobenius Theorem \cite{horn1990matrix} that 
$C$ has a unique eigenvalue at one, while all other eigenvalues are strictly inside the unit circle. 
We normalize the entries of the eigenvector that is associated with the eigenvalue at one to add up to one and refer to it as the Perron eigenvector of $C$. 
All its entries will be strictly positive. 
We  we will show in Sec. \ref{ssec:bias-analysis} and App. \ref{App:AppendixB} 
that the values for $\{ \tau_k \}$ turn out to be determined by this Perron eigenvector.  
\begin{figure}[htb]
  \centering
  \includegraphics[width=.4\linewidth]{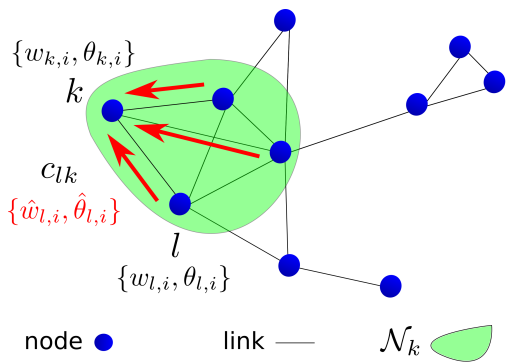}
  \caption{Example of the networks considered in this paper. 
  			Every node can only communicate with its neighbors
  			(shadowed area).}
  \label{fig:network}
\end{figure}

Iterations \eqref{eq:odgtd-adapt-theta}--\eqref{eq:odgtd-combine-w} constitute the proposed 
diffusion off-policy GTD algorithm,
which we remark is a fully distributed algorithm because the combination step is taken only over $\N_k$.

\begin{algorithm}
\caption{Diffusion off-policy GTD algorithm. 
This procedure runs in parallel at every node $k$.}
\label{alg:dogtd}
\vspace{4pt}
\textbf{Inputs:}
Target $\pi$ and behavior $\phi_k$ policies,
neighborhood $\mc{N}_k$,
weights $\{ c_{lk}, l=1,\ldots,N \}$,
and step-sizes $\mu_\theta, \mu_w$\\
\textbf{Initialize} estimates $\theta_{k,0}, \: w_{k,0}$ \\
\textbf{for} every time-step $i=1$ to $T$ \textbf{do}
\begin{itemize}
\item [] Take action $a_k(i) \sim \phi_k( \cdot | x_{k,i})$ 
\item [] Observe feature vector $ x_{k,i+1}$ and reward $r_k(i+1)$
\item [] Perform local adaptation steps \eqref{eq:odgtd-adapt-theta} and \eqref{eq:odgtd-adapt-w} 
\item [] Combine in-neighborhood estimates into $\theta_{k,i+1} , w_{k,i+1}$ using \eqref{eq:odgtd-combine-theta} and \eqref{eq:odgtd-combine-w}
\end{itemize}
\textbf{end for} \\
\textbf{Return:} $w_{k, T+1}$
\vspace{4pt}
\end{algorithm}

\section{Performance Analysis} 
\label{sec:mse_analysis}

In this section we analyze the existence and uniqueness of the optimal solution to the multi-agent learning problem \eqref{eq:global-dual-problem}. 
We extend the energy conservation arguments of  \cite{Cattivelli2010,Chen2012a,Chen2013a_distributed,SayedChapter2012}
to perform a mean-square-error (MSE) analysis of the diffusion GTD algorithm \eqref{eq:odgtd-adapt-theta}--\eqref{eq:odgtd-combine-w}
and provide convergence guarantees under sufficiently small step-sizes.
We also obtain closed form expressions of the mean-square-deviation (MSD) and analyze the bias of the algorithm.
We will rely on some reasonable conditions on the data, as explained next.

\subsection{Data model}
\label{sec:mse:data}

To begin with, we model the quantities appearing in 
\eqref{eq:odgtd-adapt-theta}--\eqref{eq:odgtd-combine-w} as instantaneous realizations of random variables, 
which we denote by using boldface notation. 
We aggregate the variables into vectors of length ${2M}$ each:
\begin{IEEEeqnarray}{rCl}
\bm{\alpha}_{k,i} 		& \defeq & 
	\left[ 
		{\begin{array}{c} 
			\bm{\theta}_{k,i}
			\\ 
			\bm{w}_{k,i} 
		\end{array}} 
	\right]
	,
\quad
	\bm{\psi}_{k,i} \defeq 
	\left[ 
		{\begin{array}{c} 
			\bm{\widehat{\theta}}_{k,i}
			\\ 
			\bm{\widehat{w}}_{k,i} 
		\end{array}} 
	\right] 
\end{IEEEeqnarray}
\begin{IEEEeqnarray}{rCl}
\bm{g}_{k,i+1} 	 & \defeq &
	\left[ 
		{\begin{array}{c} 
			- \eta \bm{x}_{k,i} \cdot \bm{\xi}_k(i) \cdot \bm{r}_k(i+1)
			\\ 
			0_M
		\end{array}} 
	\right]
\label{eq:aggregared-parameter}
\end{IEEEeqnarray}
where we are now writing $\mu_w \defeq \mu$ and $\mu_\theta  \defeq \eta  \mu_w$,
such that $\eta>0$ is the step-size ratio between the two adaptation steps.
We further introduce the following ${2M \times 2M}$ coefficient matrix:
\begin{IEEEeqnarray}{rCl}
\label{eq:aggregated-coefficient-matrix}
\bm{G}_{k,i+1} &\defeq &
	\left[
		{\begin{array}{cc}
		 		\eta \bm{x}_{k,i} \bm{x}_{k,i}^\T	 \bm{\xi}_k(i)	& 	\eta \bm{x}_{k,i} \bm{\delta}_{k,i+1}^\T \bm{\xi}_k(i)
\\
		 		- \bm{\delta}_{k,i+1}\bm{x}_{k,i}^\T \bm{\xi}_k(i)	& 				0_{M \times M}
	 	\end{array} } 
	 \right]
\;\;
\end{IEEEeqnarray}
Then, the diffusion algorithm \eqref{eq:odgtd-adapt-theta}--\eqref{eq:odgtd-combine-w}  
with stochastic variables can be expressed as
\begin{IEEEeqnarray}{rCl}
\IEEEnosubnumber
        \bm{\psi}_{k,i+1} 	& = & 	\bm{\alpha}_{k,i}
                                -
                                \mu
                                	\left( 
                                		\bm{G}_{k,i+1} \bm{\alpha}_{k,i} + \bm{g}_{k,i+1}
                                	\right)          
\IEEEyessubnumber		
\label{eq:adaptation-DSA}
\\
		\bm{\alpha}_{k,i+1} 	& = & 	\sum_{l \in \N_k} c_{lk} \bm{\psi}_{l,i+1}
\IEEEyessubnumber		
\label{eq:combination-DSA}
\end{IEEEeqnarray}
We assume the following conditions for \eqref{eq:adaptation-DSA}--\eqref{eq:combination-DSA}:
\begin{assumption} \label{assumption:iid_samples}
The state transitions $\{ ( s_k(i), s_k(i+1) ) \}$ visited by each agent $k$
are i.i.d. samples, with initial states $\{ s_k(i) \}$ drawn from the stationary distribution $d^{\phi_k}$. 
\end{assumption}
\begin{assumption} \label{assumption:non_singular} 
There is some positive probability that every state is visited by at least one agent,
thus $D^{\overline{\phi}}$ in \eqref{eq:global-stationary-distribution} is positive-definite.
\end{assumption}
\begin{assumption} \label{assumption:bounded_features_and_rewards} 
The feature matrix $X$ and the expected reward signal $r^\pi$ are bounded from below and from above.
\end{assumption}
Given the sequence of states visited by each agent 
$\{ s_k(1), s_k(2), \ldots, s_k(i), \ldots \}	$,
the segments that start and end at the same state are independent of one another.
When the Markov chain that defines these state transitions has short mixing time, these segments tend to be short (see, e.g., \cite{LevinPeresWilmer2006}).
Assumption \ref{assumption:iid_samples} approximates these independent segments with sequences of just one step.
This is a customary approximation (see, e.g., \cite{Borkar99theode,Sutton2008,Sutton2009})
that simplifies the analysis because the tuples  $\{ \bm{x}_{k,i},  \bm{a}_k(i), \bm{x}_{k,i+1}, \bm{r}_k(i) \}$ become i.i.d. samples, 
rendering $\bm{G}_{k,i+1}$ and $\bm{g}_{k,i+1}$ independent of $\bm{\alpha}_{k,i}$.

Assumption \ref{assumption:non_singular} refers to a property of the network.
For a single-agent algorithm, the agent should visit every state with positive probability;
otherwise it may not be able to approach the value function.
Here, we impose the milder condition that every state must be visited by {\em at least} one agent.

Assumption \ref{assumption:bounded_features_and_rewards} holds for most practical implementations,
and will be used in the stability analysis.

\subsection{Existence and uniqueness of solution}
\label{sec:existence-uniqueness-solution}

Solving the aggregated dual problem \eqref{eq:global-dual-problem}
is equivalent to finding the saddle-points $\{w^o,\theta^o\}$ of the global Lagrangian \eqref{eq:global-Lagrangian}.
A saddle-point of the Lagrangian must satisfy \cite{boyd2004convex}:
\begin{IEEEeqnarray}{rCl}
	L(\theta^o, w^o)	= 	\min_\theta \max_w L(\theta, w) 	= 	\max_w \min_\theta L(\theta, w)
\quad
\label{eq:saddle-point-conditions-definition}
\end{IEEEeqnarray}
These conditions are equivalent to the following system of linear equations:
\begin{IEEEeqnarray}{rcl}
	\nabla_{\theta} L(\theta, w)  	& = & 	X^\T D^{\overline{\phi}} 
												\left( 
													X \theta 
													- r^\pi 
													+ (I_S - \gamma P^\pi) X w
												\right)   
\label{eq:global-problem-dual-expanded}
									 =
											0_M 
\quad\;\;\;
\\
\label{eq:global-problem-primal-expanded}
	\nabla_{w} L(\theta, w)  		& = & 	X^\T ( I_S - \gamma P^\pi )^\T D^{\overline{\phi}} X \theta = 0_M
\end{IEEEeqnarray}
To find the saddle-point $\{\theta^o, w^o \}$, 
we solve for $\theta$ in \eqref{eq:global-problem-primal-expanded} first.
Since Assumption \ref{assumption:non_singular} establishes that $D^{\overline{\phi}}$ has full-rank,
we recall from \eqref{eq:normal-equations}--\eqref{eq:optimal-w} that
$X^\T D^{\overline{\phi}} (\gamma P^\pi - I_S) X$ is invertible 
and, hence,
$
	\theta^o		=		0_M
$. 
Then, substituting $\theta^o$ into \eqref{eq:global-problem-dual-expanded} yields:
\begin{IEEEeqnarray}{rCl}
\label{eq:dsa-optimal-parameter}
	w^o	& = &	\left(
						X^\T 
						D^{\overline{\phi}}
						\left( 
								I_S - \gamma P^\pi
						\right) 
						X	
				\right)^{-1}
				X^\T 
				D^{\overline{\phi}}
				r^\pi
\end{IEEEeqnarray}
Equation \eqref{eq:dsa-optimal-parameter} therefore illustrates one clear benefit of cooperation.
If the behavior policy of some agent prevents him from exploring the entire state-space, 
then some of the entries of its corresponding $d^{\phi_k}$ will be zero and the agent may be unable to estimate the value vector on its own.
Nevertheless, as long as any other agent in the network can visit these unexplored states, 
the matrix $D^{\overline{\phi}}$ will be positive-definite, 
guaranteeing the existence and uniqueness of a solution $w^o$.

We remark that the off-policy solution $w^o$ in \eqref{eq:dsa-optimal-parameter}
is in fact an approximation to the {\em on-policy} solution that the agents wish to predict,
which is given by \eqref{eq:optimal-w} when $D \defeq D^\pi$:
\begin{IEEEeqnarray}{rCl}
\label{eq:on-policy-solution}
	w^\pi	& = &	\left(
						X^\T 
						D^{\pi}
						\left( 
								I_S  - \gamma P^\pi
						\right) 
						X	
				\right)^{-1}
				X^\T 
				D^{\pi}
				r^\pi
\end{IEEEeqnarray}
That is, 
the obtained solution \eqref{eq:dsa-optimal-parameter} is still an approximation of \eqref{eq:on-policy-solution} 
because $D^{\bar{\phi}}$ is not necessarily the same as $D^\pi$.
However, it is interesting to realize that, by using diffusion strategies, the agents can estimate 
the exact on-policy solution if the scalars $\{\tau_k\}$ could be set to satisfy 
\begin{IEEEeqnarray}{rCl}
	\sum_{k=1}^N \tau_k 	d^{\phi_k} 		& = &		d^{\pi}		 \quad \Leftrightarrow 	\quad w^o = w^\pi
\label{eq:optimal-combination-weights}
\end{IEEEeqnarray}
%
%
%

In the next subsections, 
we analyze the conditions that allow diffusion GTD to converge to \eqref{eq:dsa-optimal-parameter}.

\subsection{Error recursion}
\label{sec:performance-analysis:error-recursion}

We introduce the following error measures, 
which measure the difference between the estimates $\{ \bm{\alpha}_{k,i}, \bm{\psi}_{k,i}\}$ at time $i$ and the optimal solution 
$\alpha^o=\mbox{\rm col}\{\theta^o,w^o\}$ for each agent $k$:
\begin{IEEEeqnarray}{rCl}
	\bm{\tilde{\psi}}_{k,i} \defeq  \alpha^o  - \bm{\psi}_{k,i}
    \label{eq:error-recursion-intermediate}
    \\
    \bm{\tilde{\alpha}}_{k,i} \defeq  \alpha^o - \bm{\alpha}_{k,i}
	\label{eq:error-recursion-parameter}
\end{IEEEeqnarray}
Then, subtracting both sides of \eqref{eq:adaptation-DSA}--\eqref{eq:combination-DSA} from $\alpha^o$,
we obtain
\begin{IEEEeqnarray}{rcl}
	\bm{\tilde{\psi}}_{k,i+1} & = & 
\label{Equ:PerformanceAnalysis:ErrorRecursion_Adapt_nodek}
                        	\left(
                                I_{2M}
                                -
                                \mu
                                \bm{G}_{k,i+1}
                            \right)
                            \bm{\tilde{\alpha}}_{k,i}
                            +
                            \mu	
                            \left(
                                \bm{G}_{k,i+1}\alpha^o + \bm{g}_{k,i+1}
                            \right)
\qquad
\end{IEEEeqnarray}
Using the fact that $c_{lk} = 0$ if $l \notin \N_k$, 
the error recursion for the combination step becomes
\begin{IEEEeqnarray}{rCl}
	\bm{\tilde{\alpha}}_{k,i}
    	=  \sum_{l \in \mc{N}_k}
        		c_{lk} \bm{\tilde{\psi}}_{l,i}
		=   \sum_{l=1}^N c_{lk} \bm{\tilde{\psi}}_{l,i}    
\label{Equ:PerformanceAnalysis:ErrorRecursion_Combine_nodek}             
\end{IEEEeqnarray}
We collect the error variables from across the network into block vectors of size $2MN$:
\begin{IEEEeqnarray}{rCl}
   \bm{\tilde{\psi}}_i     & \defeq &    \col\{
        										\bm{\tilde{\psi}}_{1,i},
                                                \ldots,
                                                \bm{\tilde{\psi}}_{N,i}
                                            \}
	\\
	\bm{\tilde{\alpha}}_i    & \defeq &     \col\{
                                                    \bm{\tilde{\alpha}}_{1,i},
                                                    \ldots,
                                                    \bm{\tilde{\alpha}}_{N,i}
                                            \}
\end{IEEEeqnarray}
Let $\mc{C}$ and $\bm{\mc{R}}_{i}$ be matrices of size $2MN \times 2MN$ defined by
\begin{IEEEeqnarray}{rCl}
	\mc{C}	& \defeq &	C \otimes I_{2M} 
\label{eq:network_C} 
\\
	\bm{\mc{R}}_{i}	 & \defeq &
                             	\diag\{
                                	\bm{G}_{1,i},
                                    \ldots,
									\bm{G}_{N,i}
                                 \}
\label{eq:network-coefficient-matrix}
\end{IEEEeqnarray}
and let $\bm{\mc{G}}_{i}$ be the matrix of size $2MN \times 2M$ defined by
\begin{IEEEeqnarray}{rCl}
	\bm{\mc{G}}_{i}  & \defeq &     \col\{
												\bm{G}_{1,i},
                                                	\ldots,
                                                    \bm{G}_{N,i}
                                            \}
\end{IEEEeqnarray}
We also introduce the vectors of length $2MN$:
\begin{IEEEeqnarray}{rCl}
	\bm{g}_{i}     & \defeq &    \col\{
                                \bm{g}_{1,i},
                         		\ldots,
                                \bm{g}_{N,i}
                            \}
\\
	\bm{n}_{i}    & \defeq &    \bm{\mc{G}}_{i} \alpha^o + \bm{g}_{i}
\end{IEEEeqnarray}
Then, the individual error recursions in \eqref{eq:error-recursion-intermediate}--\eqref{eq:error-recursion-parameter} lead to the 
following network recursion:
\begin{IEEEeqnarray}{rCl}
	\bm{\tilde{\alpha}}_{i+1}   =   \mc{C}^\T
    									\big(
                                        	I_{2MN} - \mu \bm{\mc{R}}_{i+1}
                                        \big)
                                        \bm{\tilde{\alpha}}_{i}
                                        +
                                        \mu
                                        \mc{C}^\T
                                        \bm{n}_{i+1}
\label{eq:ErrorRecursion_final}
\end{IEEEeqnarray}
This recursion shows how the error dynamics evolves over the network over time.


\subsection{Convergence in the mean}
\label{Sec:PerformanceAnalysis:MeanConvergence}


Introduce the following expected values for each agent: 
\begin{IEEEeqnarray}{rCl}
\label{eq:expected_G_k}
	G_k     	& 	\defeq	&	\E \bm{G}_{k,i} 
	=
			\left[
				{\begin{array}{cc}
			 		\eta X^\T D^{\phi_k} X 							& 		\eta X^\T D^{\phi_k} (I_S - \gamma P^{\pi})X   		\\
		 			- X^\T (I_S - \gamma P^{\pi})^\T D^{\phi_k} X 	& 				0_{M \times M}   	\\
			 	\end{array} } 
			 \right]
\\
\label{eq:expected_g_k}
	g_k     	& 	\defeq	&	\E \bm{g}_{k,i} 
				=
					\left[
							{\begin{array}{c}
						 		- \eta X^\T D^{\phi_k} r^\pi	\\
					 			0_M   	\\
						 	\end{array} } 
					 \right]
\end{IEEEeqnarray}
Since Assumption \ref{assumption:iid_samples} implies that the variables 
$\bm{\mc{R}}_{i+1}$ and $\bm{\tilde{\alpha}}_{i}$ are independent of each other,
then by taking expectations of both sides of \eqref{eq:ErrorRecursion_final}
we obtain
    \begin{IEEEeqnarray}{rCl}
        \label{Equ:PerformanceAnalysis:MeanErrorRecursion}
        \E {\bm{\tilde\alpha}_{i+1}} =   \mc{C}^\T
                                            \big(
                                                    I_{2MN} - \mu {\mc{R}}
                                            \big)
                                            \E \bm{\tilde{\alpha}}_{i} 
                                            + \mu \mc{C}^\T (\mc{G}\alpha^o + g)
\quad
    \end{IEEEeqnarray}
where 
\begin{IEEEeqnarray}{rCl}
\label{eq:expected_R}
	\mc{R}     	& 	\defeq	&	\E\bm{\mc{R}}_{i} 
					= 
								\diag\{
                                	G_1,
                                    \ldots,
									G_N
                                 \}
\\
\label{eq:expected_network_mcG}
	\mc{G} 		&	=	& 		\E \bm{\mc{G}}_i 
					= 			\col\{
									G_1,
                              		\ldots,
                                    G_N
								\}
\\
\label{eq:expected_network_g}
	g  			&	= 	&		\E \bm{g}_i	
					= 			\col\{
									g_1,
                              		\ldots,
                                    g_N
								\}
\end{IEEEeqnarray}
Therefore, 
the convergence of \eqref{Equ:PerformanceAnalysis:MeanErrorRecursion} is guaranteed 
when the matrix $ \mc{C}^\T ( I_{2MN} - \mu {\mc{R}} )$ is stable.
\begin{theorem}[Mean convergence]
	\label{Thm:MeanStability}
	For the data model of Section \ref{sec:mse:data},
	there exists small enough step-sizes, 
	say $0 < \mu <\mu^o$ 
	(for some $\mu^o>0$ given by \eqref{eq:bound-step-size} in Appendix \ref{App:AppendixA}), 
	such that 
	the matrix $ \mc{C}^\T ( I_{2MN} - \mu {\mc{R}} )$ is stable and,
	therefore, the mean-error recursion \eqref{Equ:PerformanceAnalysis:MeanErrorRecursion}
	is stable for every agent $k=1, \ldots, N$ 
	and converges to the bias value given by
	\begin{IEEEeqnarray}{rCl}
		\tilde{\alpha}_{\infty} 
			& \defeq &
				\lim_{ i \rightarrow \infty } \E \bm{\tilde{\alpha}}_i 
	\label{eq:bias-closed-form-expression}
			=
						\left(
							I_{2MN} 
							- \mc{C}^\T 
							\left(
								I_{2MN} - \mu \mc{R}
							\right)
						\right)^{-1}
						\mu 
						\mc{C}^\T 
							\left(
								\mc{G}\alpha^o + g
							\right)								
	\nonumber 
	\end{IEEEeqnarray} 
\end{theorem}
\begin{IEEEproof}
See Appendix \ref{App:AppendixA}.
\end{IEEEproof}
As it is explained in Appendix \ref{App:AppendixA},
the value $\mu^o$ only depends on the inputs of the algorithm, namely, data-samples (state-features and transition rewards),
the weighted-topology matrix $C$, 
the cost-weights $\{\tau_k \}$,
and the step-size ratio parameter $\eta$.


\subsection{Mean-square stability}

Although the error vector converges in the mean, 
we still need to ensure that it has bounded fluctuations around its fixed point value. 
To do so, we study the evolution and steady-state value of the variance 
$
	\E \| \bm{\tilde{\alpha}}_{i} \|^2
$.
By computing the weighted squared Euclidean (semi)norm of both sides of
\eqref{eq:ErrorRecursion_final}---using an arbitrary positive 
(semi)definite weighting matrix $\Sigma$ that we are free to choose---and applying the expectation operator, 
we obtain the following variance relation:
\begin{IEEEeqnarray}{rCl}
	\E\| \bm{\tilde{\alpha}}_{i+1}\|_{\Sigma}^2 
                        &=&  \E
                        	 \left \| 
                        	 	\bm{\tilde{\alpha}}_{i}
                        	 \right \|_{\Sigma'}^2
                                +
                                2b_{\Sigma}^\T 
                                	\; \E \bm{\tilde{\alpha}}_{i}
                                +
                                \mathrm{Tr}
                                \left(
                                	\mu^2
                                    \Sigma \mc{C}^\T \mc{R}_n \mc{C}
                                \right)
\qquad
\label{Equ:PerformanceAnalysis:WeightedMSE_relation}
\end{IEEEeqnarray}
where
\begin{IEEEeqnarray}{rCl}
        \Sigma'  
        			& \defeq & 
                        (I_{2MN} - \mu \mc{R}^\T) \mc{C}\Sigma\mc{C}^\T (I_{2MN} 
                        	- \mu \mc{R})
							+ 
                        			\mu^2
                        			\E
                        			\left[
                        					(\bm{\mc{R}}_{i+1} - \mc{R})^\T
                        					 	\mc{C}\Sigma\mc{C}^\T
                        					 	(\bm{\mc{R}}_{i+1} 
                        					 - \mc{R})
                        			\right]
\quad
\label{Equ:PerformanceAnalysis:Sigma_prime}
%
%
\\
        b_\Sigma        & \defeq & \mu 
        						\E
                                \left[
                                    \left(
                                    	I_{2MN} - \mu \bm{\mc{R}}_{i+1}^\T 
                                    \right)
                                    \mc{C}
                                    \Sigma
                                    \mc{C}^\T
                                    \bm{n}_{i+1}
                                \right]
\label{Equ:PerformanceAnalysis:b_Sigma}
\\
        \mc{R}_n		& \defeq & 	\E\left[\bm{n}_i \bm{n}_i^\T\right]
        				=	\E \left[ 
        							(\bm{\mc{G}}_{i} \alpha^o + \bm{g}_{i})
        							(\bm{\mc{G}}_{i} \alpha^o + \bm{g}_{i})^\T 
    	    					\right]
\label{Equ:PerformanceAnalysis:R_v}
\end{IEEEeqnarray}
Let $\sigma=\mathrm{vec}(\Sigma)$.
Using the Kronecker product property 
$\mathrm{vec}(Y \Sigma Z)=(Z^\T \otimes Y) \mathrm{vec}(\Sigma)$
\cite{Sayed2008}, 
we can vectorize $\Sigma'$ in \eqref{Equ:PerformanceAnalysis:Sigma_prime}
and find that its vector form is related to $\Sigma$ via the following \emph{linear} relation:
$\sigma' \defeq \mathrm{vec}(\Sigma') = \mc{F} \sigma$, where the matrix $\mc{F}$ is given by
\begin{IEEEeqnarray}{rCl}
\label{eq:matrix-F-state-space-model}
		\mc{F}	& \defeq &	
						\left(
								\left(I_{2MN} - \mu \mc{R}^\T \right)
								\mc{C}
						\right) 
						\otimes 
						\left(
								\left( I_{2MN} - \mu \mc{R}^\T \right)
								\mc{C}
						\right)
					+ 
						\mu^2
						\E
						\left[
							\left( 
								(\bm{\mc{R}}_{i+1}^\T-\mc{R}^\T) \mc{C} 
							\right)
							\otimes
							\left(
								(\bm{\mc{R}}_{i+1}^\T-\mc{R}^\T) \mc{C}
							\right)
						\right]
\notag \\
\end{IEEEeqnarray}
Furthermore, using the property $\mathrm{Tr}(\Sigma Y) = ( \mathrm{vec}[Y^\T] )^\T \sigma$, we can rewrite 
\eqref{Equ:PerformanceAnalysis:WeightedMSE_relation} as:
\begin{IEEEeqnarray}{rCl}
	\label{Equ:PerformanceAnalysis:Weighted_VarianceRelation_final}
	\E\| \bm{\tilde{\alpha}}_{i+1} \|_{\sigma}^2		= 		
														\E
														\|
															\bm{\tilde{\alpha}}_{i}
														\|_{\mc{F}\sigma}^2 
														+ 
														2\sigma^\T  
														\mc{U} \cdot
														\E \bm{\tilde{\alpha}}_{i}
														+
														h^\T
														\sigma
\end{IEEEeqnarray}
where 
\begin{IEEEeqnarray}{rCl}
		\label{Equ:PerformanceAnalysis:H_def}
		\mc{U}	& \defeq &	\mu 
							\E \big[
									\left(
										\mc{C}^\T 
										\bm{n}_{i+1}
									\right)
									\otimes
									\left(
										\mc{C}^\T 
											\left( 
												I_{2MN} - \mu \bm{\mc{R}}_{i+1} 
											\right) 
									\right)
								\big]
\\
		\label{Equ:PerformanceAnalysis:r_def}
		h		& \defeq & 	
						\mu^2
						\mathrm{vec}
						\left[ 
							\mc{C}^\T
							\mc{R}_n
							\mc{C}
						\right]
\end{IEEEeqnarray}
In \eqref{Equ:PerformanceAnalysis:Weighted_VarianceRelation_final} we are using the notation $\|x\|^2_{\sigma}$ to represent $\|x\|^2_{\Sigma}$.  
Note that 
\eqref{Equ:PerformanceAnalysis:Weighted_VarianceRelation_final}
is not a true recursion because the weighting matrices corresponding to $\sigma$ and $\mc{F}\sigma$ are different.
Moreover, recursion \eqref{Equ:PerformanceAnalysis:Weighted_VarianceRelation_final} is coupled with the mean-error recursion \eqref{Equ:PerformanceAnalysis:MeanErrorRecursion}.
To study the convergence of \eqref{Equ:PerformanceAnalysis:Weighted_VarianceRelation_final}
we will expand it into a state-space model following \cite{Sayed2008,al-naffouri_transient_2003}.
Let $L \defeq 2MN$ and let $p(x)$ denote the characteristic polynomial
of the $L^2 \times L^2$ matrix $\mc{F}$, 
given by
\begin{equation}
	p(x)	
			\defeq 
				\mathrm{det}(x I - \mc{F})
			= 
				x^{L^2} + p_{L^2-1} x^{L^2-1} + \ldots + p_0
\end{equation}
By the Cayley-Hamilton Theorem \cite{Sayed2008}, 
we know that every matrix satisfies its characteristic equation (i.e., $p(\mc{F})=0$), 
so that
\begin{IEEEeqnarray}{rCl}
	\mc{F}^{L^2}	=	-p_0 I_{L^2} - p_1 \mc{F} - \ldots - p_{L^2-1} \mc{F}^{L^2-1}
\end{IEEEeqnarray}
Replacing $\sigma$ in \eqref{Equ:PerformanceAnalysis:Weighted_VarianceRelation_final} by $\mc{F}^j \sigma$, $j=0,\ldots,L^2-1$, 
we can derive the following state-space model:
	\begin{IEEEeqnarray}{rcl}
		\underbrace{
		\left[\begin{IEEEeqnarraybox*}[\mysmallarraydecl]
		[c]{c}
			\E\| \bm{\tilde{\alpha}}_{i+1} \|_\sigma^2	\\
			\E\| \bm{\tilde{\alpha}}_{i+1} \|_{\mc{F}\sigma}^2	\\
			\vdots								\\
			\E\| \bm{\tilde{\alpha}}_{i+1} \|_{\mc{F}^{L^2-1}\sigma}^2
		\end{IEEEeqnarraybox*}\right]}_{\mc{W}_{i+1}}
		&&	=	
				\underbrace{
				\left[\begin{IEEEeqnarraybox*}[\mysmallarraydecl]
				[c]{c,c,c,c,c}
					0	&	1	&	0	&	\cdots	&	0		\\
					0	&	0	&	1	&	\cdots	&	0		\\
					\vdots&		&		&	\ddots	&	0		\\
					0	&	0	&	0	&	\cdots	&	1		\\
					-p_0&-p_1	&-p_2	&	\cdots	&	-p_{L^2-1}
				\end{IEEEeqnarraybox*}\right]}_{\mc{T}}
				\underbrace{
				\left[\begin{IEEEeqnarraybox*}[\mysmallarraydecl]
				[c]{c}
					\E\| \bm{\tilde{\alpha}}_{i}\|_\sigma^2	\\
					\E\| \bm{\tilde{\alpha}}_{i}\|_{\mc{F}\sigma}^2	\\
					\vdots								\\
					\E\| \bm{\tilde{\alpha}}_{i}\|_{\mc{F}^{L^2-1}\sigma}^2
				\end{IEEEeqnarraybox*}\right]}_{\mc{W}_{i}}
		\label{Equ:PerformanceAnalysis:MSE_recursion}
				+
				2
				\underbrace{
					\left[
						\begin{IEEEeqnarraybox*}[\mysmallarraydecl]
						[c]{c}
							\sigma^\T \mc{U}		\\
							\sigma^\T\mc{F}\mc{U}	\\
							\vdots					\\
							\sigma^\T\mc{F}^{L^2-1}\mc{U}
						\end{IEEEeqnarraybox*}
					\right]
				}_{\mc{Q}}
				\E \bm{\tilde{\alpha}}_{i}
				+
				\underbrace{
				\left[\begin{IEEEeqnarraybox*}[\mysmallarraydecl]
				[c]{c}
					h^\T \sigma				\\
					h^\T	\mc{F} \sigma		\\
					\vdots					\\
					h^\T	\mc{F}^{L^2-1}\sigma
				\end{IEEEeqnarraybox*}\right]}_{\mc{Y}}
				\IEEEeqnarraynumspace
	\end{IEEEeqnarray}
We combine \eqref{Equ:PerformanceAnalysis:MSE_recursion} 
with the mean-recursion \eqref{Equ:PerformanceAnalysis:MeanErrorRecursion} and rewrite them more compactly as:
\begin{IEEEeqnarray}{rCl}
		\label{Equ:PerformanceAnalysis:Joint_Recursion}
		\begin{bmatrix}
			\mc{W}_{i+1}	\\
			\E \bm{\tilde{\alpha}}_{i+1}
		\end{bmatrix}
				&=&		\begin{bmatrix}
							\mc{T}	&	2\mc{Q}		\\
							0		&	\mc{C}^\T(I_{2MN} - \mu \mc{R})
						\end{bmatrix}
						\begin{bmatrix}
							\mc{W}_{i}	\\
							\E \bm{\tilde{\alpha}}_{i}
						\end{bmatrix}
						+
						\begin{bmatrix}
							\mc{Y}		\\
							\mc{C}^\T 
							\mc{G}\alpha^o + g	
						\end{bmatrix}
		\IEEEeqnarraynumspace
\end{IEEEeqnarray}

\begin{theorem}[Mean-square stability]
	\label{Thm:MeanSquareStability}
	Assume the step-size parameter $\mu$ is sufficiently small 
	so that terms that depend on higher-order powers of $\mu$ can be ignored.  
	Then, for the data model of Section \ref{sec:mse:data},
	there exists $0 < \mu_{\mathrm{MS}}^o \le \mu^o $ 
	(for $\mu^o$ used in Theorem \ref{Thm:MeanStability} and given by \eqref{eq:bound-step-size} in Appendix \ref{App:AppendixA}), 
	such that when $0 < \mu < \mu_{\mathrm{MS}}^o$,
	the variance recursion \eqref{Equ:PerformanceAnalysis:Joint_Recursion} is mean-square stable. 
\end{theorem}
\begin{IEEEproof}
Observe that the stability of the joint recursion
\eqref{Equ:PerformanceAnalysis:Joint_Recursion} is equivalent to
the stability of the matrices $\mc{T}$ and $\mc{C}^\T(I_{2MN}-\mu \mc{R})$,
which is further equivalent to the following conditions on
their spectral radii:
\begin{IEEEeqnarray}{rCl}
		\label{Equ:PerformanceAnalysis:MeanSquareStability_Condition_original}
			\rho\left(\mc{C}^\T(I_{2MN}-\mu \mc{R})\right)<1				
			, \quad
			\rho\left(\mc{T}\right)<1			
\end{IEEEeqnarray} 
The first condition is the same mean-stability condition that was discussed in
Theorem \ref{Thm:MeanStability}. 
For the second condition,
we note from \eqref{Equ:PerformanceAnalysis:MSE_recursion} 
that $\mc{T}$ is in companion form, and it is known that its eigenvalues are the roots of $p(x)$, 
which are also the eigenvalues of $\mc{F}$. 
Therefore, a necessary and sufficient condition for the stability of $\mc{T}$ is the stability of the matrix $\mc{F}$.
When the step-sizes are small enough, 
the last term in \eqref{eq:matrix-F-state-space-model} can be ignored since it depends on $\mu^2$ and we can write 
\begin{IEEEeqnarray}{rCl}
		\mc{F}	&\approx&
						\left(
								\mc{C}^\T\left(I_{2MN}- \mu \mc{R}\right)
						\right)^\T
						\otimes 
						\left(
								\mc{C}^\T\left(I_{2MN}- \mu \mc{R}\right)
						\right)^\T 
				\IEEEeqnarraynumspace
\label{eq:approx_F}
\end{IEEEeqnarray}
which is stable if
$\mc{C}^\T\left(I_{2MN}-\mu \mc{R}\right)$ is stable. 
\end{IEEEproof}
We remark that $0 < \mu_{\mathrm{MS}}^o \le \mu^o $ is chosen to dismiss higher-order powers of $\mu$, 
and that $\mu^o$ (see Appendix \ref{App:AppendixA}) only depends on the inputs of the algorithm 
(i.e., data-samples, 
the weighted-topology matrix $C$, 
the weights $\{\tau_k \}$ and the parameter $\eta$).

\subsection{Mean-square performance}

Taking the limit of both sides of \eqref{Equ:PerformanceAnalysis:Weighted_VarianceRelation_final} we obtain:
\begin{IEEEeqnarray}{rCl}
		\lim_{i \rightarrow \infty} 
			\E\| \bm{\tilde{\alpha}}_{i+1}\|_{\sigma}^2
				&=&	\lim_{i \rightarrow \infty}
					\E\| \bm{\tilde{\alpha}}_{i}\|_{\mc{F}\sigma}^2
					+
					2\sigma^\T
					\mc{U}
					\lim_{i \rightarrow\infty}
					\E \bm{\tilde{\alpha}}_{i}
					+
					h^\T\sigma
\label{eq:steady-state-variance-recursion}
\end{IEEEeqnarray}
Theorem \ref{Thm:MeanStability} guarantees that
$\lim_{i \rightarrow\infty} \E \bm{\tilde{\alpha}}_{i} = \tilde{\alpha}_\infty$,
so the steady-state variance recursion in \eqref{eq:steady-state-variance-recursion} leads to
\begin{IEEEeqnarray}{rCl}
\label{Equ:PerformanceAnalysis:SteadyStateMSE_general}
	\lim_{i \rightarrow \infty} \E\| \bm{\tilde{\alpha}}_{i}\|_{\sigma}^2
		& = &	
			q^\T(I-\mc{F})^{-1}\sigma
\end{IEEEeqnarray}
where
$
	q 		\defeq 		h + 2 \mc{U} \tilde{\alpha}_\infty
$. 
Result \eqref{Equ:PerformanceAnalysis:SteadyStateMSE_general} is useful because it allows us to derive several performance metrics through the
proper selection of the free weighting parameter vector $\sigma$ 
(or, equivalently, the parameter matrix $\Sigma$).
For example, the network mean-square-deviation (MSD) 
is defined as the average of the MSD of all the agents in the network:
\begin{IEEEeqnarray}{rCl}
\mathsf{MSD}^{\text{network}} 	& \triangleq &
									\lim_{i \rightarrow \infty} 
										\frac{1}{N} \sum_{k=1}^N 
											\E \| \bm{\tilde{\alpha}}_{k,i} \|^2 
								=
									\lim_{i \rightarrow \infty} 
										\E\| \bm{\tilde{\alpha}}_{i}\|_{\frac{1}{N} I_{2MN}}^2
\end{IEEEeqnarray}
Choosing the weighting matrix in \eqref{Equ:PerformanceAnalysis:SteadyStateMSE_general} as $\Sigma=I_{2MN}/N$, we get:
\begin{IEEEeqnarray}{rCl}
\boxed{
	\mathsf{MSD}^{\text{network}} 	= 	\frac{1}{N} 
											q^\T (I-\mc{F})^{-1} \text{vec}(I_{2MN})
	}
\end{IEEEeqnarray}
We can also obtain the MSD of any particular node $k$, 
as
\begin{IEEEeqnarray}{rCl}
\mathsf{MSD}_k 	& \triangleq &	\lim_{i \rightarrow \infty} 
								\E \| \bm{\tilde{\alpha}}_{i} \|^2_{\mathcal{J}_k}
\end{IEEEeqnarray}
where $\mathcal{J}_k$ is a block-diagonal matrix 
of $N$ blocks of size $2M \times 2M$, such that all blocks in the diagonal are zero
except for block $k$ which is the identity matrix.
Following the same procedure as with the network MSD we obtain
\begin{IEEEeqnarray}{rCl}
\boxed{
	\mathsf{MSD}_k 	=	 	q^\T (I-\mc{F})^{-1} \text{vec} (\mc{J}_k)
	}
\end{IEEEeqnarray}


\subsection{Bias analysis}
\label{ssec:bias-analysis}

We showed in \eqref{eq:dsa-optimal-parameter}  
that, under Assumption \ref{assumption:non_singular}, 
there exists a unique solution $\alpha^o$
for the global optimization problem \eqref{eq:global-dual-problem}.
On the other hand, the error recursion \eqref{Equ:PerformanceAnalysis:MeanErrorRecursion}
converges in the mean-square sense to some bias value $\tilde{\alpha}_{\infty}$.
Now, we examine under which conditions $\tilde{\alpha}_{\infty}$ is small
when the step-size is small.
The analysis of the bias value $\tilde{\alpha}_{\infty}$ in \eqref{eq:bias-closed-form-expression} is similar to the examination developed in \cite[Theorem 3]{Chen2013a_distributed} 
for multi-objective optimization. 
The main difference lies in the fact that we do not assume 
the matrix $\mc{R}$ in \eqref{eq:expected_R} to be symmetric.

\begin{theorem}[Bias at small step-size]
\label{theorem:bias}
	Consider the data model of Section \ref{sec:mse:data},
	where the combination matrix $C$ is primitive left-stochastic 
	(it satisfies \eqref{eq:Comb_matrix_1_2_3}--\eqref{eq:Comb_matrix_4}).
	Suppose the Perron eigenvector that corresponds to the eigenvalue of $C$ at one is 	equal to the vector of weights $\{ \tau_k \}$ in the global problem \eqref{eq:global-dual-problem}
	(i.e., $\tau^\T C = \tau^\T$ and $\tau^\T \mathds{1}_{N} = 1$).
	Assume further that the step-size $\mu$ 
	is sufficiently small to ensure mean-square stability. Then, it holds that 
\begin{IEEEeqnarray}{rCl}
	\boxed{
		\tilde{\alpha}_\infty 	= 	O(\mu)
	}
	\label{eq:bias-theorem}
\end{IEEEeqnarray}
\end{theorem}
\begin{IEEEproof}
See Appendix \ref{App:AppendixB}.
\end{IEEEproof}

We remark that the bias in \eqref{eq:bias-theorem} comes from agents following different behavior policies,
which means that they are solving different optimization problems, 
with different minimizer each.
When they use diffusion strategies, 
the combination step pulls them toward the global solution.
Nevertheless, the adaptation step pushes each agent towards the minimizer of its individual cost function.
Note, however, that if all agents followed the same behavior policy,
their individual optimization problems would be identical,
therefore, both the adaptation and the combination steps would pull them toward the global solution
and their fixed-point estimates would be unbiased with respect to the solution of the global optimization problem \eqref{eq:global-dual-problem},
as stated in \cite{Valcarcel2013}.
More formally, $G_k \defeq \bar{G}$ and $g_k \defeq \bar{g}$ would be the same for every agent $k$,
and the saddle-point conditions of the Lagrangian of the global problem \eqref{eq:global-Lagrangian}
would not depend on the combination weights $\{\tau_k\}$:
\begin{equation*}
	\sum_{k=1}^N \tau_k  ( \bar{G} \alpha^o + \bar{g}) = \bar{G} \alpha^o + \bar{g} = 0_{2M}
\label{eq:global-problem-with-same-policy}
\end{equation*}
Therefore, if all the agents followed the same behavioral policy,
then $\mc{G} \alpha^o + g = 0_{2MN}$ and, from \eqref{eq:bias-closed-form-expression},
we would conclude that $\tilde{\alpha}_\infty = 0$.


\section{Simulations}
\label{sec:simulations}

\begin{figure}[htb]
  \centering
  \includegraphics[width=.5\linewidth]{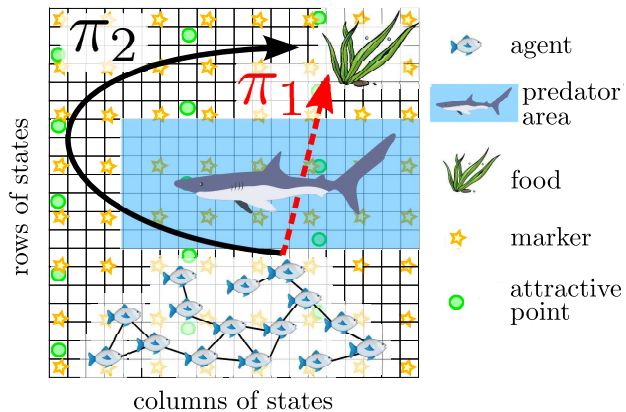}
  \caption{Problem representation.
  			Network of $15$ randomly connected agents.
  			States are numbered from left to right and from down to top, 
  			so that the bottom-left corner corresponds with $s=1$, and the top-right corner with $s=S=400$.
  			The agents are attracted towards the center of a region each (green circle) at states $\{395, 368, \ldots, 21 \}$.
  			The predator's area is the rectangle $5 \le {\rm column} \le 20, 8 \le {\rm row} \le 15$.
  			The reward at every time step is $-1$, 
  			except when crossing the predator's area where it becomes $-15$, and
  			when eating the food when it rises to $+20$.
    			The $64$ features are obtained with $8$ Gaussian basis functions per row and per column
    			and with standard deviation $0.005$, 
    			equidistantly located at positions (yellow stars) that are computed by taking the grid as a continuous unit-square.
  			}
  \label{fig:2d-world}
\end{figure}

Consider a group of animals foraging in a $2$D-world (see Figure \ref{fig:2d-world}).
The group forms a network of $N=15$ agents with arbitrarily connected topology
and neighborhood size $|\mc{N}|_k$ varying between $2$ and $9$.
The weights of the links (i.e., the $c_{lk}$ elements of $C$)
are obtained independently by each node following an \textit{averaging rule} \cite{Blondel2005,SayedChapter2012},
such that equal weight is given to any member of the neighborhood, including itself
(i.e., $c_{lk} = 1 / |\mc{N}_k|, l \in \mc{N}_k$). 
Note this rule leads to a left (rather than doubly) stochastic combination matrix $C$ 
that satisfies \eqref{eq:Comb_matrix_1_2_3}--\eqref{eq:Comb_matrix_4}.
We assume that the combination matrix $C$---and, hence, the network topology---remains fixed.

The world is a discrete, bounded square with $20$ rows and $20$ columns,
which amounts to $S=400$ states.  
Each agent self-localizes itself in the grid by sensing a Gaussian radial basis function of its distance to $M=64$ fixed markers 
(i.e., each of these values is a feature). 
The agents move in four possible directions, 
namely, $\Ac = \{$north, south, east, west$\}$.
At every time step, the agents move and consume some energy (i.e., they receive some negative reward).
In the north-east corner of the world, there is food, which the agents understand as positive reward.
However, there is a large area below the food with a predator that is harmful to go through, 
so agents receive large negative reward if they visit these states (see caption of Figure \ref{fig:2d-world} to see the exact numerical values).

Since the agents are getting negative reward at every time step (because of energy consumption) 
they want to know how to reach the food, 
while losing as less energy as possible.
A natural policy, denoted $\pi_1$, could be to go straight to the food with high ($0.8$) probability and low ($0.2$) probability of going in another direction, 
but then the agents would face the harmful predator and the total expected reward may be low.
Thus, we say that $\pi_1$ is a myopic policy.
Another more insightful policy, denoted $\pi_2$, could be to take a detour and avoid the predator's area with very high ($0.95$) probability. 
Nevertheless, if the detour takes too long, then the agents would consume too much energy and it may not be worth trying. 
In order to evaluate which policy is better (myopic $\pi_1$ or detour $\pi_2$),
the agents have to learn the value vector of each candidate-policy from samples. 
If the agents were learning {\em on-policy}, they would have to follow one candidate policy for long enough 
so they could apply stochastic optimization over the samples, 
then they would have to start again but following the other candidate policy. 
In other words, on-policy learning does not allow to reuse samples while evaluating different policies.
The benefit of the {\em off-policy} formulation is that the agents can evaluate several target policies in parallel from a single data stream. 

We consider the case in which the agents are territorial and tend to settle in different regions each.
In other words, 
the behavior policies of the agents $\{ \phi_k \}$ are all different and constrain exploration to some regions of the state-space.
At every time-step, each agent is attracted to the state at the center of its territory with $0.8$ probability, 
and it moves in a different direction with $0.2$ probability.
Since the agents only have samples of state-transitions in their respective territories, 
it is difficult for them to predict the value vector for each of the target policies ($\pi_1$ and $\pi_2$).
However, since they sample complementary regions of the state-space, 
they can collaborate, applying diffusion strategies, 
to learn the value vector of the two target policies and evaluate which one is better.

\begin{figure}[htb]
  \centering
  \includegraphics[width=.9\linewidth]{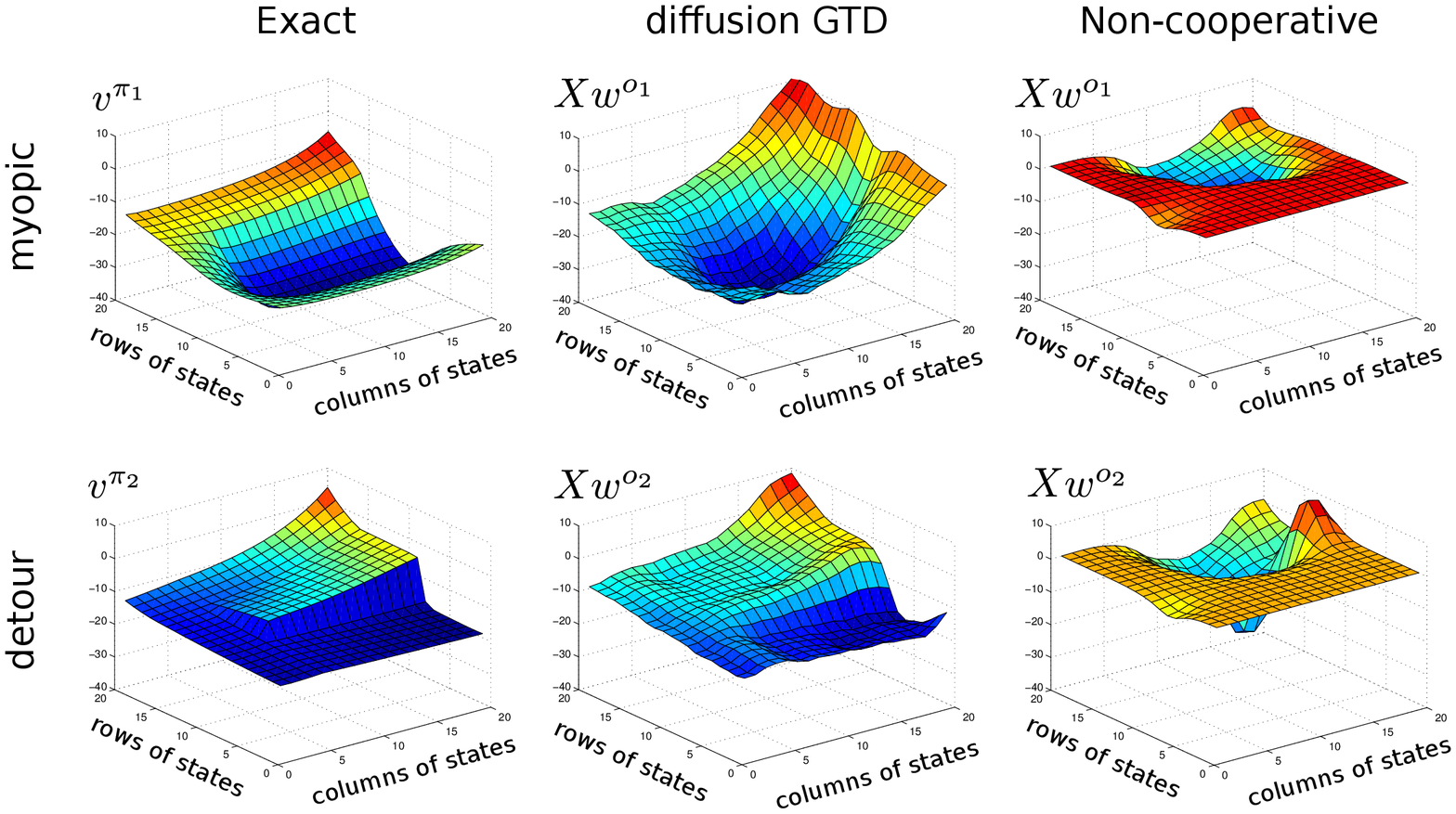}
  \caption{Value vectors for myopic (top, $\pi_1$) and detour (bottom, $\pi_2$) policies for each state of the grid.
  			The left column shows the exact value vectors computed using Eq. \eqref{eq:linear-Bellman-equation}.
  			The myopic policy shows a valley because the agents have many chances of entering into the predator's area, 
			so the expected reward will be negative. 
			For the detour policy, the agent has little chance of entering into the predator's region, 
			so the expected value will be near zero in this area.
  			The middle column shows the off-policy, linearly approximated and cooperatively estimated value vector obtained with our diffusion GTD algorithm 
  			when the agents follow the non-exploratory policies (they move towards an attractive point each).
  			The agents are able to reproduce the main features of the surfaces (valleys, hills and flat areas).
  			The right column shows the same value vectors but estimated with no cooperation (for a node that did not diverge).
  			}
  \label{fig:value_vectors}
\end{figure}

\begin{figure}[htb]
  \centering
  \includegraphics[width=0.9\linewidth]{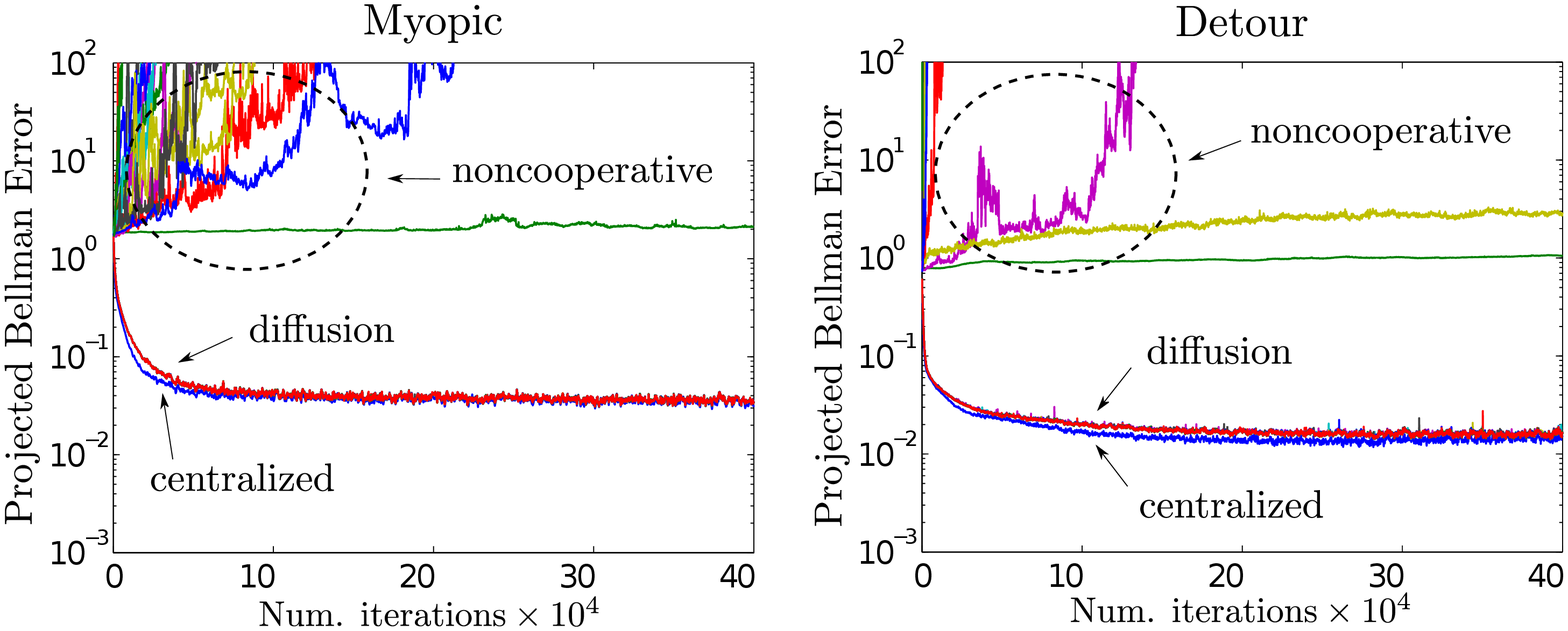}
  \caption{Projected Bellman error (i.e., $J_{\rm PB}$) 
  	given by Eq. \eqref{eq:projected-Bellman-error} for centralized, 
  	diffusion and non-cooperative solutions.
  	Results are obtained with the following parameters.
  	We set the long-term reward to $\gamma = 0.95$,
	which we have found to make $v^{\pi_1}$ very different from $v^{\pi_2}$ for the specific values of $\pi_1$, $\pi_2$ and rewards
	used during the experiments.
	For both cooperative and non-cooperative nodes,
	the step-size and step-size ratio are $\mu=1 \times 10^{-3}$ and $\eta=0.1$,
	which adequately trade bias and convergence-rate to illustrate the behavior of the algorithm.
	We assume every agent takes one sample at every time-step.
	Since a central station would have $N$ times more samples than the rest of the agents,
	we set its step-size equal to $\mu/N$.
	Results are averaged over $50$ independent experiments.
  			}
  \label{fig:mspbe}
\end{figure}

Figure \ref{fig:value_vectors} 
shows\footnote{Code available at http://gaps.ssr.upm.es/images/sergio/tsp-coop-pred-20131004.zip} the exact value vector for the myopic and detour policies,
as well as its cooperative--using the proposed diffusion GTD algorithm--and non-cooperative approximation for one agent, 
under the considered constrained-exploration off-policy multi-agent setting.
Figure \ref{fig:mspbe} shows the learning curve of the algorithm. 
Since the agents have only samples from small portions of the state-space, 
the non-cooperative algorithm may diverge. 
On the other hand,  
when the agents cooperate 
(i.e., communicate their estimates to their neighbors),
the diffusion algorithm allows them to benefit from the experience from other agents in the network, 
and they approach the same solution as a centralized architecture 
(i.e., with a fusion center that gathers all the samples from every node) 
would achieve but more efficiently, by communicating only within neighborhoods. 

\section{Conclusion}

Diffusion GTD maintains  the efficiency of the single-agent GTD2 \cite{Sutton2009}, 
with linear complexity in both computation time and memory footprint.
With diffusion GTD, 
the agents learn directly from samples
(without any apriori knowledge of the environment) 
and cooperate to improve the stability and accuracy of their prediction.
We remark that cooperation is fully distributed with
communications only within each agent's neighborhood;
neither fusion-center, nor multi-hop communications are required.

We provided conditions that guarantee convergence
of the proposed diffusion GTD 
and derived performance bounds for sufficiently small step-sizes.
Although our analysis assumes stationarity,
constant step-sizes are a desirable feature for an adaptive network,
since it allows the network to learn continuously, and to track concept drifts in the data.

\begin{appendices}

\section{Proof of Theorem \ref{Thm:MeanStability}} \label{App:AppendixA}

To study the spectrum of $ \mc{C}^\T 	\left( I_{2MN} - \mu {\mc{R}} \right)$, 
we express the combination coefficient matrix
in its Jordan canonical form:
\begin{IEEEeqnarray}{rCl}
\label{eq:jordan_canonical_C}
	C^\T      =  Y_C J_C Y_C^{-1}
\end{IEEEeqnarray}
Using the property $(U \otimes V) (Y \otimes Z) = (U Y) \otimes (V Z)$, 
we obtain 
\begin{IEEEeqnarray}{rCl}
\mc{C}^\T 
	( 
		I_{2MN} 
		- \mu 
		 {\mc{R}} 
	)
&
			=
&
				(Y_C \otimes I_{2M}) 
				(J_C \otimes I_{2M}) 
				(Y_C^{-1} \otimes I_{2M})
				\big( 
					I_{2MN} - \mu {\mc{R}} 
				\big)
\nonumber
\\
			& 
			= 
			&
				(Y_C \otimes I_{2M}) 
				(J_C \otimes I_{2M}) 
				\big(
					I_{2MN} 
					-
					\mu 
					(Y_C^{-1} \otimes I_{2M}) 
					{\mc{R}} 
					(Y_C \otimes I_{2M})
				\big)
				(Y_C^{-1} \otimes I_{2M}) 
\qquad
\label{eq:network-stability-matrix}
\end{IEEEeqnarray}
so that, by similarity,
\begin{IEEEeqnarray}{rCl}
\lambda 
	\left( 
		\mc{C}^\T 
			\left( 
				I_{2MN} - \mu {\mc{R}} 
			\right) 
	\right)
	&=&
\lambda 
	\left( 
		(J_C \otimes I_{2M}) 
		\left(
			I_{2MN} 
			- \mu \mc{E}
		\right) 
	\right)
\qquad
\label{eq:similar-mat}
\end{IEEEeqnarray}
where
\begin{IEEEeqnarray}{rCl}
\label{eq:mat_E}
\mc{E} 		\defeq		 (Y_C^{-1} \otimes I_{2M}) \mc{R} (Y_C \otimes I_{2M}) 
\end{IEEEeqnarray}
As stated in Section \ref{sec:multi-agent}, conditions \eqref{eq:Comb_matrix_1_2_3}--\eqref{eq:Comb_matrix_4} 
ensure that $C^\T$ is a primitive right-stochastic matrix.
Hence, from the Perron-Frobenius Theorem \cite{horn1990matrix}, 
the Jordan canonical form of $C$ can be expressed as
\begin{IEEEeqnarray}{rCl}
\label{eq:split_canonical}
	J_C 			=	 	\diag \left\lbrace 1, J_C^0 \right\rbrace
\end{IEEEeqnarray}
where all the eigenvalues of $J_C^0$ are strictly inside the unit circle.
Moreover, since $C^\T$ is right-stochastic it has one right-eigenvector of all ones 
associated with its unit eigenvalue,
and its corresponding left eigenvector, $p$, has positive entries (i.e., $p_k > 0, \: 1 \le k \le N$):
\begin{IEEEeqnarray}{rCl}
\label{eq:left-eigenvector-C}
	C^\T \mathds{1}_N = \mathds{1}_N
	, \;\;
	p^\T C^\T = p^\T
	, \;\;
	\mathds{1}_N^\T p = 1 
\end{IEEEeqnarray}
We therefore decompose 
\begin{IEEEeqnarray}{rCl}
\label{eq:eigenvectors-C}
	Y_C^{-1} 	=		\col{ \left\lbrace 
									p^\T,
										\:
										Y_C^l	
									\right\rbrace }
,\quad
	Y_C = 	\left[
				\mathds{1}_N \; 
				Y_C^r
			\:
			\right]
\end{IEEEeqnarray}
and partition $\mc{E}$ as
\begin{IEEEeqnarray}{rCl}
\mc{E} 		& = &
	\left[
		\begin{array}{cc}
			\bar{G} & \mc{E}_{12} \\
			\mc{E}_{21} & \mc{E}_{22}
		\end{array}
	\right]
\label{eq:partitioned_E}
\\
\bar{G}
		& \defeq & 	
			\left( 
				p^\T 
				\otimes 
				I_{2M}
			\right)
			\mc{R}
			\left(
				\mathds{1}_N \otimes I_{2M}
			\right)
		= 
			\sum_{k=1}^N p_k G_k 	
\label{eq:G_combined}
\\
\mc{E}_{12}	& \defeq & 	
			\left(
				p^\T 
				\otimes 
				I_{2M}
			\right)
			\mc{R}
			\left(
				Y_C^r \otimes I_{2M}
			\right)
\\
\mc{E}_{21}	& \defeq & 	
			\left(
				Y_C^l
				\otimes 
				I_{2M}
			\right)
			\mc{R}
			\left(
				\mathds{1}_N \otimes I_{2M}
			\right)
\label{eq:definition-E_21}
\\
\mc{E}_{22}	& \defeq & 	
			\left(
				Y_C^l
				\otimes 
				I_{2M}
			\right)
			\mc{R}
			\left(
				Y_C^r \otimes I_{2M}
			\right)
\label{eq:E22}
\end{IEEEeqnarray}
Introduce the following shorthand in \eqref{eq:similar-mat}:
\begin{IEEEeqnarray}{rCl}
\label{eq:S}
\mc{S} \defeq 	(J_C \otimes I_{2M}) \left( I_{2MN} - \mu \mc{E} \right)
\end{IEEEeqnarray}
Then, expanding \eqref{eq:split_canonical} and \eqref{eq:partitioned_E}-\eqref{eq:E22} into \eqref{eq:S} 
we have
\begin{IEEEeqnarray}{rCl}
\label{eq:similar-stable-mat}
\mc{S}
	& = &
		\left[
			\begin{array}{cc}
				I_{2M} - \mu \bar{G}
							&		-\mu \mc{E}_{12}					\\
				-\mu \left( J_C^0 \otimes I_{2M} \right) \mc{E}_{21}		
							&		\left( J_C^0 \otimes I_{2M} \right) 
									\left( I_L - \mu \mc{E}_{22} \right)
			\end{array}
		\right]
\quad\;\;\:
\end{IEEEeqnarray}
where $L \defeq  2M(N-1)$.
Using the same technique proposed in  \cite{Zhao2014,Chen2014}, we appeal to eigenvalue perturbation analysis 
to examine the spectral radius of \eqref{eq:similar-stable-mat}.
We introduce the $N\times N$ diagonal matrix 
$
	\Omega_N^\epsilon \defeq \diag \{ \epsilon, \epsilon^2, \epsilon^3, \ldots, \epsilon^N \}
$
with parameter $\epsilon > 0$.
Let 
$
{J_{\bar{G}}} = Y_{\bar{G}}^{-1} \bar{G} Y_{\bar{G}}
$
be the Jordan canonical form of $\bar{G}$ and introduce the similarity transformation
\begin{IEEEeqnarray}{rCl}
\label{eq:similarity-transformation-phi}
	\Phi = 
		\left[
			\begin{array}{cc}
				Y_{\bar{G}}	\Omega_{2M}^{\epsilon}		& 	0_{2M \times L}			\\
				0_{L \times 2M}							&	\frac{\sqrt{\mu}}{\sigma} \Omega_{N-1}^{\beta} \otimes I_{2M} 
			\end{array}
		\right]
\end{IEEEeqnarray}
with parameters $\epsilon$, $\beta$, and $\sigma$. 
We apply the similarity transformation \eqref{eq:similarity-transformation-phi} to $\mc{S}$: 
\begin{IEEEeqnarray}{c}
	\Phi^{-1}
	\mc{S}
	\Phi
=
\left[
		\begin{array}{cc}
			I_{2M} - \mu J_{\bar{G}}^\epsilon		
							& 	- \frac{\mu\sqrt{\mu}}{\sigma} \mc{S}_{12}	\\
			- \sigma \sqrt{\mu} \: \mc{S}_{21}
							&	J_C^{0 \beta} \otimes I_{2M} - \mu \mc{S}_{22}
		\end{array}
\right]
\label{eq:similarity-transformation}
\end{IEEEeqnarray}
where 
\begin{IEEEeqnarray}{rCl}
	\mc{S}_{12}
		& \defeq &
			(\Omega_{2M}^\epsilon)^{-1} 
			Y_{\bar{G}}^{-1} 
			\mc{E}_{12} 
			\left(
				\Omega_{N-1}^{\beta} \otimes I_{2M}
			\right)
\\
	\mc{S}_{21}
		& \defeq &
			\left(
				\left(
					(\Omega_{N-1}^\beta)^{-1} J_C^{0} 
				\right) 
					\otimes I_{2M}
			\right) 
			\mc{E}_{21} 
			Y_{\bar{G}}
			\Omega_{2M}^{\epsilon}
\label{eq:definition-S_21}
\\
	\mc{S}_{22}
		& \defeq &
				\left(
					\left(
						(\Omega_{N-1}^\beta)^{-1} J_C^{0} 
					\right)
					\otimes I_{2M}
				\right)
				\mc{E}_{22}
				\left(
					\Omega_{N-1}^\beta \otimes I_{2M}
				\right)
\quad\;
\end{IEEEeqnarray}
and
$J_{\bar{G}}^\epsilon$ 
and 
$J_C^{0 \beta}$
have the same form as the (upper triangular) Jordan canonical forms $J_{\bar{G}}$ and $J_C^{0}$, 
except that the unit entries are replaced by $\epsilon$ and $\beta$, respectively.
By applying Gerschgorin theorem \cite{horn1990matrix} to \eqref{eq:similarity-transformation},
we can identify the regions where the eigenvalues of $\mc{S}$ should lie:
\begin{IEEEeqnarray}{rCl}
	\big| 
		\lambda(
		\mc{S}) 
		- 
		\big( 
			1 
		- 
		\mu 	\lambda_{m}(\bar{G}) 
	   \big)
	\big|
	& 
	\le 
	&
		\mu \epsilon 
		+
		\frac{\mu \sqrt{\mu}}{\sigma}
		\sum_{q =1}^L 
			\left| 
				\left[ 
					\mc{S}_{12}
				\right]_{mq} 
			\right|
\label{eq:Gershgoring-disk-1}
\\
	\big| 
		\lambda(
		\mc{S}) 
		- 
		\big( 
			\lambda_{k+1}(C) 
			- 
			\mu  
			\left[ 
				\mc{S}_{22}
			\right]_{mm} 
	   \big)
	\big|
	& \le &
		\beta 
		+
		\sigma
		\sqrt{\mu}
		\sum_{q =1}^{2M} 
			\left| 
				\left[ 
					\mc{S}_{21}
				\right]_{mq} 
			\right|
		+
		\mu 
		\sum_{\substack{q =1 \\ q \ne m}}^{L}  
			\left| 
				\left[ 
					\mc{S}_{22}
				\right]_{mq} 
			\right| 
\qquad
\label{eq:Gershgoring-disk-2}
\end{IEEEeqnarray}
where $[ \cdot ]_{mq}$ stands for the element at row $m$ and column $q$ of a matrix.
Although we use the same subscript $m$ in both equations, note that $1 \le m \le 2M$ in \eqref{eq:Gershgoring-disk-1} while 
$1 \le m \le L$ in \eqref{eq:Gershgoring-disk-2}; 
in addition, recall that $C$ is an $N \times N$ matrix, 
hence, we use subscript $k = {\rm ceil} \{ m /(2M) \}$, 
where ${\rm ceil} \{ \cdot \}$ rounds a real number to its nearest greater or equal integer.

We are looking for sufficient conditions that guarantee that the mean recursion \eqref{Equ:PerformanceAnalysis:MeanErrorRecursion} converges for small step-size.
Recall from \eqref{eq:similar-mat} that \eqref{Equ:PerformanceAnalysis:MeanErrorRecursion} converges when $|\lambda(\mc{S})| < 1$. 
Let us solve for $|\lambda(\mc{S})|$ in \eqref{eq:Gershgoring-disk-1} first. 
Since $|z| - |y| \le |z - y| $, we obtain
\begin{IEEEeqnarray}{rCl}
\left | 
	\lambda(\mc{S}) 
\right | 
	- 
	\left | 
		1- \mu \lambda_m(\bar{G}) 
	\right |
	&	
	\le
	&
		\left | 
			\lambda(\mc{S})
			- 
			( 1- \mu \lambda_m(\bar{G}) )
		\right |
	\le
		\mu \epsilon 
		+
		\frac{\mu \sqrt{\mu}}{\sigma}
		\chi_{(12)}
\label{eq:substract-absolute-value}
\end{IEEEeqnarray}
where 
$
	\chi_{(12)}
	\defeq
		\sum_{q =1}^L 
			\left| 
				\left[ 
					\mc{S}_{12}
				\right]_{mq} 
			\right|
$.
Therefore,
\begin{IEEEeqnarray}{rCl}
	| \lambda(\mc{S}) | 
	&	
	\le
	&
		\mu \epsilon 
		+
		\frac{\mu \sqrt{\mu}}{\sigma}
		\chi_{(12)}
		+
		\left| 
			1- \mu \lambda_m(\bar{G}) 
		\right|
\label{eq:Stability-condition-1}
\end{IEEEeqnarray}
Since $\mc{S}$, $\bar{G}$ and $C$ are not generally guaranteed to be symmetric, their eigenvalues may be complex.
Using the fact that $1-z \le  \left(  1 - \frac{1}{2}z  \right)^2$ for $z \in \Re$,
we obtain
\begin{IEEEeqnarray}{rCl}
	\left| 
		1- \mu \lambda_m(\bar{G}) 
	\right|
	&
	\le
	&
		1
		-
		\mu {\rm Re}\{\lambda_m(\bar{G})\}
		+
		\frac{\mu^2}{2} 
			\left| \lambda_m(\bar{G})
			\right|^2
\quad
\label{eq:substract-absolute-value-2}
\end{IEEEeqnarray}
where ${\rm Re}\{\cdot\}$ denotes the real part of a complex number.
Combining \eqref{eq:Stability-condition-1} and \eqref{eq:substract-absolute-value-2}, 
the stability condition implied by \eqref{eq:Gershgoring-disk-1} requires finding small step-sizes $\mu$ such that
\begin{IEEEeqnarray}{rCl}
	\mu 
	\frac{|\lambda_m(\bar{G})|^2}{2} 
	+	
	\sqrt{\mu} \frac{\chi_{(12)}}{\sigma} 
	+
	\epsilon
	-
	{\rm Re}\{\lambda_m(\bar{G})\}
&
	<
&
		0
\quad
\end{IEEEeqnarray}
which leads to
\begin{IEEEeqnarray}{rCl}
	0
&&
	<
		\mu
	<
\label{eq:step-size-condition-1}
		\left(
			\frac
			{
				- \frac{\chi_{(12)}}{\sigma} 
				+
				\sqrt
				{
					\left(
						\frac{\chi_{(12)}}{\sigma}
					\right)^2
					+
					2 |\lambda_m(\bar{G})|^2
					(
						{\rm Re}\{\lambda_m(\bar{G})\} 
						- 
						\epsilon
					)
				}
			}
			{
				|\lambda_m(\bar{G})|^2
			}
		\right)^2
\end{IEEEeqnarray}
where,
in order to guarantee that the term inside the square root in the right side of \eqref{eq:step-size-condition-1} is positive,
we choose
\begin{IEEEeqnarray}{rCl}
	0 < \epsilon < \min_{1\le m \le 2M} {\rm Re} \{ \lambda_m(\bar{G}) \}
\end{IEEEeqnarray}
We now show that ${\rm Re} \{ \lambda_m(\bar{G}) \}$ is always positive.
If we transform $\bar{G}$ into a similar matrix:
\begin{IEEEeqnarray}{rCl}
	\bar{G}_{\sqrt{\eta}}
		& \defeq &
			\scriptsize{
				\left[
					{\begin{array}{cc}
					 		I_M 				& 		0_{M \times M}		 \\
			 				0_{M \times M}				& 		\sqrt{\eta}I_M\\
				 	\end{array} } 
				\right]
				 	 \bar{G}
				\left[
					{\begin{array}{cc}
			 				I_M 					& 		0_{M \times M}  			 \\
					 		0_{M \times M}			& 		\frac{1}{\sqrt{\eta}} I_M\\
				 	\end{array} } 
				 \right]
			}
		=
			\left[
			{\scriptsize
				{\begin{array}{cc}
			 		\sqrt{\eta} X^\T D^{\overline{\phi}} X 					& 		X^\T D^{\overline{\phi}} (I_S - \gamma P^{\pi})X   		\\
		 			- X(I_S - \gamma P^{\pi})^\T D^{\overline{\phi}} X^\T 	& 				0_{M \times M}   	\\
			 	\end{array} } 
			 }
			 \right]
\qquad\;
\label{eq:G-similar}
\end{IEEEeqnarray}
and use \cite[Theorem 3.6]{Benzi2005} on $\bar{G}_{\sqrt{\eta}}$, 
we can establish that ${\rm Re} \{ \lambda_m( \bar{G} ) \} >0$. 

Now, we solve for $|\lambda(\mc{S})|$ from \eqref{eq:Gershgoring-disk-2}.
Let us abbreviate the sums in the right side of \eqref{eq:Gershgoring-disk-2} as
\begin{IEEEeqnarray}{rCl}
	\chi_{(21)} 
	&
	\defeq
	&
		\sum_{q =1}^{2M} 
			\left| 
				[ \mc{S}_{21} ]_{mq} 
			\right|
,\quad
\chi_{(22)} 
	\defeq
		\sum_{\substack{q =1 \\ q \ne m}}^{L}  
			\left| 
				[ \mc{S}_{22} ]_{mq} 
			\right| 
\quad
\label{eq:define-sums-2}
\end{IEEEeqnarray}
In a manner similar to \eqref{eq:substract-absolute-value}, we have
\begin{IEEEeqnarray}{rCl}
	| \lambda(\mc{S}) |
	- 
	\left|
		\lambda_{k+1}(C) 
			- 
			\mu  
			[ \mc{S}_{22} ]_{mm} 
	\right|
&
	\le
&
		\beta 
		+
		\sigma
		\sqrt{\mu}
		\chi_{(21)} 
		+
		\mu \chi_{(22)} 
\label{eq:derivation-condition-disk-2}
\end{IEEEeqnarray}
Using \eqref{eq:derivation-condition-disk-2} 
and the fact $|z-y| \le |z| + |y|$ yields the following condition on $\mu$ for stability:
\begin{IEEEeqnarray}{rCl}
\sqrt{\mu}
	\big(
		\sqrt{\mu} 
			\left( 
				\chi_{(22)}  
				+ 
				| [ \mc{S}_{22} ]_{mm} |
			\right)
&&
		+ 
		\sigma \chi_{(21)} 
	\big)
	< 
		1 
		-
		\left| 
			\lambda_{k+1}(C) 
		\right|
		-
		\beta 
\label{eq:condition-step-size-derivation-2}
\end{IEEEeqnarray}
The following conditions on the step-size are jointly sufficient to satisfy \eqref{eq:condition-step-size-derivation-2}:
\begin{IEEEeqnarray}{rCl}
	0	<	\sqrt{\mu} 	< 1 - | \lambda_{k+1}(C) | - \beta 
\label{eq:condition-step-size-derivation-2-a}
\\
	0 	<	
			\sqrt{\mu} 
					< 
						\frac
						{
							1
							- 
							\sigma \chi_{(21)}
						}
						{
							\chi_{(22)} 
							+ 
							| [ \mc{S}_{22} ]_{mm} |
						}
\label{eq:condition-step-size-derivation-2-b}
\end{IEEEeqnarray}
From the Perron-Frobenius, we know that $\lambda_{k+1}(C) < 1$, for $1 \le k \le N-1$.
Moreover, Assumption \ref{assumption:bounded_features_and_rewards} guarantees that any element of $\mc{S}$ is bounded from below and above.
Therefore, there exist parameters
$
0  <  \beta  < 1 - |\lambda_{k+1}(C)|
$
and
$ 
0 < 	\sigma	< 1 / \chi_{(21)}
$
that make the right side of \eqref{eq:condition-step-size-derivation-2-a} and \eqref{eq:condition-step-size-derivation-2-b} positive, respectively.
Hence,
we can square both inequalities and obtain the following conditions:
\begin{IEEEeqnarray}{rCl}
	0	
&
	<
&
		\mu 	
			< 
				\left( 
					1 - | \lambda_{k+1}(C) | - \beta 
				\right)^2
\label{eq:step-size-condition-2-a}
\\
	0 	
&
	<
&
		\mu 
			< 
				\left(
					\frac
					{
						1
						- 
						\sigma \chi_{(21)}
					}
					{
						\chi_{(22)} 
						+ 
						| [ \mc{S}_{22} ]_{mm} |
					}
				\right)^2
\label{eq:step-size-condition-2-b}
\end{IEEEeqnarray}
Let us bound \eqref{eq:step-size-condition-1}, \eqref{eq:step-size-condition-2-a}
and \eqref{eq:step-size-condition-2-b} by
\begin{IEEEeqnarray}{rCl}
&&
	\mu^o 
		= \min
		\Bigg\lbrace		
		\left( 
						1 - | \lambda_{k+1}(C) | - \beta 
					\right)^2
					, 
		\bigg(
						\frac
						{
							1
							- 
							\sigma \chi_{(21)}
						}
						{
							\chi_{(22)} 
							+ 
							| [ \mc{S}_{22} ]_{jj} |
						}
		\bigg)^2
,
\notag
\\
\label{eq:bound-step-size}
&&		
		\left(
			\frac
			{
				- \frac{\chi_{(12)}}{\sigma} 
				+
				\sqrt
				{
					\left(
						\frac{\chi_{(12)}}{\sigma}
					\right)^2
					+
					2 |\lambda_m(\bar{G})|^2
					(
						{\rm Re}\{\lambda_m(\bar{G})\} 
						- 
						\epsilon
					)
				}
			}
			{
				|\lambda_m(\bar{G})|^2
			}
		\right)^2
	\Bigg\rbrace
\end{IEEEeqnarray}
for $1 \le m \le 2M$, $1 \le k \le N-1$
and $1 \le j \le L$.
We conclude that if the step-size $0 < \mu < \mu^o$,
then diffusion GTD is mean-stable.

As a final remark, note that $\mu^o$ depends on the eigenvalues of $\bar{G}$, 
the eigenvalues of the weighted-topology matrix $C$, 
and the constructed matrix $\cal{S}$ in \eqref{eq:similar-stable-mat}
(note that $\sigma, \epsilon$ and $\beta$ are similarity parameters, 
and the terms $\chi_{(22)}$ and $\chi_{(22)}$ are defined in \eqref{eq:define-sums-2} simply as a short hand of sums of the elements in $\cal{S}$).
Recall that $\bar{G}$ is given by \eqref{eq:G_combined} as the weighted sum of the individual $G_k$,
which only depend on the data samples, 
the importance weights and the step-size ratio parameter $\eta$.
Finally, recall that $\mc{S}$ depends on the Jordan canonical form of $C$ 
and $\mc{R}$, where the latter is defined in \eqref{eq:expected_R} from the individual $G_k$.
Thus, all the terms involved in $\mu^o$ are input data to the algorithm.


\section{Proof of Theorem \ref{theorem:bias}} \label{App:AppendixB}


We follow an argument similar to \cite{Chen2013a_distributed,zhao_performance_2012}. 
It suffices to show that 
$
	\lim_{\mu \rightarrow 0} \frac{\| \tilde{\alpha}_{\infty} \|}{\mu} = \varepsilon_o
$,
where $\varepsilon_o$ is a constant independent of $\mu$.
Substituting \eqref{eq:network-stability-matrix}, \eqref{eq:mat_E} and \eqref{eq:similar-stable-mat} into \eqref{eq:bias-closed-form-expression} yields
\begin{IEEEeqnarray}{rCl}
\label{eq:bias-Kronecker}
\tilde{\alpha}_{\infty}
	& = &
			\mu 
			\left( 
				Y_C \otimes I_{2M} 
			\right)
			\left( 
				I_{2MN} - \mc{S}
			\right)^{-1}	
			( Y_C^{-1} \otimes I_{2M} )
			\mc{C}^\T 
				\left(
					\mc{G}\alpha^o + g
				\right)	
\end{IEEEeqnarray}
Expanding 
\eqref{eq:network_C},
\eqref{eq:jordan_canonical_C} and 
\eqref{eq:eigenvectors-C}
into \eqref{eq:bias-Kronecker} leads to
\begin{IEEEeqnarray}{rCl}
\tilde{\alpha}_{\infty}
	& = &
		\mu
		( Y_C \otimes I_{2M} )
		( I_{2MN} - \mc{S} )^{-1}
		( J_C \otimes I_{2M} )
			\left[
				\begin{array}{c}
					\left( p^\T \otimes I_{2M} \right) \left( \mc{G}\alpha^o + g \right)	\\
					\left( Y_C^l \otimes I_{2M} \right) \left( \mc{G}\alpha^o + g \right)						
				\end{array}
			\right]
\label{eq:bias-block-matrices}
\end{IEEEeqnarray}
From now on, assume that the weights used in \eqref{eq:global-dual-problem} in defining the global cost are the entries of the Perron eigenvector of $C$,
i.e.,
$
	p \defeq \tau
$.
Then, the first row of the last term in \eqref{eq:bias-block-matrices} stands for the saddle-point conditions of the Lagrangian of the global problem \eqref{eq:global-Lagrangian}:
\begin{IEEEeqnarray}{rCl}
	( \tau^\T 
&&
		\otimes I_{2M} ) ( \mc{G}\alpha^o + g )	
	=
		\sum_{k=1}^N \tau_k 
			\left( 
				G_k \alpha^o + g_k 
			\right)
	= 
		\left[
		{\scriptsize
			{\begin{array}{cc}
		 		\eta X^\T D^{\overline{\phi}} ( X	\theta^o 	+  (I_S - \gamma P^\pi ) X w^o  - r^\pi )
\\
	 			- X^\T (I_S - \gamma P^\pi )^\T D^{\overline{\phi}} X \theta^o
		 	\end{array} } 
		}
		\right]					 
	= 
		0_{2M}
\qquad
\label{eq:Global_Lagrangian_in_vector_form}
\end{IEEEeqnarray}
Therefore, expanding $J_C$ into \eqref{eq:bias-block-matrices} yields
\begin{IEEEeqnarray}{rCl}
\tilde{\alpha}_{\infty}
	& = &
		\mu
		( Y_C \otimes I_{2M} )
		( I_{2MN} - \mc{S} )^{-1}
			\left[
				\begin{array}{c}
					0_{2M}	\\
					( J_C^0 \otimes I_{2M} ) ( Y_C^l \otimes I_{2M} ) \left( \mc{G}\alpha^o + g \right)						
				\end{array}
			\right]
\label{eq:bias-block-matrices-shorter-form}
\end{IEEEeqnarray}
From \eqref{eq:similar-stable-mat}, 
we know that the upper-left block of $I_{2MN} - \mc{S}$ is given by $\mu \bar{G}$.
By using \cite[Theorem 3.6]{Benzi2005}, 
we establish that the similar matrix $\bar{G}_{\sqrt{\eta}}$ in \eqref{eq:G-similar} is invertible.
Therefore, $\bar{G}$ is also invertible so we can use the following relation
\begin{IEEEeqnarray}{rCl}
\left[
{\scriptsize
	\begin{array}{cc}
		Z_{11}		&		Z_{12}	\\
		Z_{21}		&		Z_{22}
	\end{array}
}
\right]^{-1}
=
\left[
{\scriptsize
	\begin{array}{cc}
		Z_{11}^{-1} + Z_{11}^{-1} Z_{12} U Z_{21} Z_{11}^{-1}		&		-Z_{11}^{-1} Z_{12} U	\\
				- U Z_{21} Z_{11}^{-1}								&			U
	\end{array}
}
\right]
\nonumber
\end{IEEEeqnarray}
where $U = (Z_{22} - Z_{21} Z_{11}^{-1} Z_{12})^{-1}$,
to write
\begin{IEEEeqnarray}{rCl}
(I_{2MN} - \mc{S})^{-1}
	= 
	\left[
			\begin{array}{cc}
				H_{11}		&		H_{12}					\\
				H_{21}		&		H_{22}
			\end{array}
	\right]
\end{IEEEeqnarray}
In this way, relation \eqref{eq:bias-block-matrices-shorter-form} simplifies to 
\begin{IEEEeqnarray}{rCl}
\tilde{\alpha}_\infty 
	& = & 
	\mu \left( Y_C \otimes I_{2M} \right)
	\left[
		\begin{array}{c}
			H_{12} 	( J_C^0 Y_C^l \otimes I_{2M} ) 	( \mc{G}\alpha^o + g ) \\
			H_{22} 	( J_C^0 Y_C^l \otimes I_{2M} ) 	( \mc{G}\alpha^o + g ) 
		\end{array}
	\right]
\qquad
\label{eq:bias-final}
\end{IEEEeqnarray}
The only terms in \eqref{eq:bias-final} that depend on $\mu$ are $H_{12}$ and $H_{22}$.
In the limit, when $\mu \rightarrow 0$ these two terms become independent of $\mu$:
\begin{IEEEeqnarray}{rCl}
\lim_{\mu \rightarrow 0} H_{12} 	& = &		- \bar{G}^{-1} \mc{E}_{12} ( I_{2MN} - J_C^0 \otimes I_{2M} )^{-1}
\\
\lim_{\mu \rightarrow 0} H_{22} 	& = & 		( I_{2MN} - J_C^0 \otimes I_{2M} )^{-1}
\end{IEEEeqnarray}
Hence, we can conclude that $\lim_{\mu \rightarrow 0} \frac{\| \tilde{\alpha}_{\infty} \|}{\mu} = \varepsilon_o$.

\end{appendices}

\bibliographystyle{IEEEtran}
\bibliography{IEEEabrv,refs}

\end{document}